# Flow based features and validation metric for machine learning reconstruction of PIV data

Ghasem Akbari[1*], Nader Montazerin[2]

[1] *Department of Mechanical Engineering, Qazvin Branch, Islamic Azad University, Qazvin, Iran. Email: g.akbari@aut.ac.ir*

[2] *Department of Mechanical Engineering, Amirkabir University of Technology, Tehran, Iran. Email: mntzrn@aut.ac.ir*

**Abstract:** Reconstruction of flow field from real sparse data by a physics-oriented approach is a current challenge for fluid scientists in the AI community. The problem includes feature recognition and implementation of AI algorithms that link data to a physical feature space in order to produce reconstructed data. The present article applies machine learning approach to study contribution of different flow-based features with practical fluid mechanics applications for reconstruction of the missing data of turbomachinery PIV measurements. Support vector regression (SVR) and multi-layer perceptron (MLP) are selected as two robust regressors capable of modelling non-linear fluid flow phenomena. The proposed flow-based features are optimally scaled and filtered to extract the best configuration. In addition to conventional data-based validation of the regressors, a metric is proposed that reflects mass conservation law as an important requirement for a physical flow reproduction. For a velocity field including 25% of clustered missing data, the reconstruction accuracy achieved by SVR in terms of $R_2$-score is as high as 0.993 for the in-plane velocity vectors in comparison with that obtained by MLP which is up to 0.981. In terms of mass conservation metric, the SVR model by $R_2$-score up to 0.96 is considerably more accurate than the MLP estimator. For extremely sparse data with a gappiness of 75%, vector and contour plots from SVR and MLP were consistent with those of the original field.

---

* **Corresponding Author;** Department of Mechanical Engineering, Qazvin Branch, Islamic Azad University, Qazvin, Iran, P.O. Box: 34185-1416.
Email: g.akbari@aut.ac.ir; Phone & Fax: +98 28 3367 0051.

**1. Introduction**

Technological race towards building reliable velocity measuring instrumentation with high temporal and spatial resolution is moving fast. Particle image velocimetry (PIV) and particle tracking velocimetry (PTV) are two expensive methods that are prone to gappy data and require careful post processing [1]. This needs reliable schemes that recompense the use of affordable low resolution equipment, with the calculation of high frequency/wave number flow structures and reconstruction of the missing data.

Velocity reconstruction with physical methods results in high resolution and is powerful against noise or data sparsity. Use of the vorticity transport equation is one such methodology that works independent of boundary conditions [2]. An 'all-in-one method' was developed for validation of PIV raw data, correction of spurious and missing data and smoothing of the outcome field [3]. The penalized least square method was also used to automatically classify the flow outliers into scattered and clustered ones [4]. These methods are computationally expensive and are only applicable where numerical pre-estimates of the velocity field are achievable or at least where boundary and initial conditions are known a priori [5, 6].

Statistical methods identify and subsequently eliminate individual vectors that depart from the ensemble of the recorded velocity flow field and eventually substitute them with the corrected vectors [1, 7-10]. In just a few examples of such methods, Nogueira et al. proposed a linear approach based on local Taylor expansions to filter and correct false vectors in PIV data [11]. This was robust and resulted in straightforward optimizations with limited accuracy. The PIV missing data was corrected in another study through element-by-element multiplication of correlation matrix calculated from adjacent PIV nodes [12]. This approach resulted in bias error reduction and subpixel accuracy enhancement. Bootstrapping is another approach that independently generates statistics for each velocity component and is capable of both outlier detection and correction of flow field [13].

Application of proper orthogonal decomposition (POD) to extract the DNS- and PIV-based modes as well as utilization of Kriging interpolation to smooth the obtained POD modes has proven to enhance reconstruction accuracy but with increased computational cost [14]. Everson and Sirovich [15] proposed a physical POD based approach combined with the least squares scheme for flow field image reconstruction from gappy data. Venturi and Karniadakis [16] developed an extended version termed gappy proper orthogonal decomposition (GPOD). It was robust and effective, while its higher computational cost in comparison with that of Everson and Sirovich approach was a drawback. The limitation of using a constant number of POD modes in traditional gappy POD methods was overcome by implementing a local adaptive criterion for detecting an optimized count of modes [17].

On outlier detection and reconstruction, Wang et al. [18] proposed a more efficient but slightly slower than conventional POD approach. This method iteratively identified and replaced clustered outliers using POD and dynamically reproduced the target velocity field. A successful algorithm for detection of outliers for single PIV frames and their velocity field reconstruction without decreasing the estimation accuracy was proposed based on a non-iterative alternative POD scheme [19]. Another gappy POD method, applied to the PIV data of high-pressure gas turbine combustor flow with low signal-to-noise ratio, adaptively decided whether a local missing data is updated in each iteration [20]. This method resulted in lower reconstruction error than the previous GPOD approaches as well as the Kriging interpolation. Application of a POD-based fusion approach was successful in recovering the missing data, especially when the flow field had dominating flow structures that were closely correlated in space [21]. Such approaches do not consider the physical characteristics of the flow and normally smooth velocity gradients or lower spatial resolutions.

Nowadays, machine learning (ML) algorithms are extensively utilized for a diverse range of fluid flow problems. Powerful capability of machine learning in forming a generalized non-linear relationship between input space and target flow quantities makes it a plausible contender for handling complicated flow data [22]. There are four different applications of ML algorithms in PIV-based flow analysis or related disciplines:

(1) *Handling raw images to improve their defects for further post-processing*: Implementation of convolutional neural network (CNN) to Remove spurious particle images and efficiently regenerate them with the same resolution works for either low- or high-particle seeding reconstructions [23]. CNN was also successfully applied by Gao et al. [24] to refine the particle reconstructions from very course initial guess and any classic algebraic PIV processing technique.

(2) *Implementation of ML approach to extract the velocity field from raw acquired PIV images*: Cai et al. [25, 26] used a deep learning estimator in terms of a convolutional neural network to extract the velocity field from a pair of PIV images. They showed similar accuracy as other state of the art methods and high efficiency toward a real-time estimation for both artificial and experimental images. Satisfactory accuracy and computational efficiency were also obtained from a shallow neural network model for PTV application [27]. Barway et al. [28] used CNN to a combination of OH-PLIF (planar laser-induced fluorescence of Hydroxyl Radical) and PIV to build a velocity field in a strong swirling combustor. They decoded the three-dimensional PIV fields from input OH-PLIF images, as a practical approach when simultaneous measurements were limited. Tombal et al. [29] applied both artificial Neural Network (ANN) and Support Vector Machine (SVM) to PIV data of particle transport in multi-phase flow. They showed that SVM as a classifier was the best for direction estimation and support vector regression (SVR) was the best for velocity approximation.

(3) *Utilization of ML algorithms to enhance temporal or spatial resolution of PIV data*: Time-resolved flow reconstruction is another recent outcome of machine learning. Implementation of long short-term memory (LSTM)-based networks and bidirectional recurrent neural networks, and their combination with POD confirmed the potential of this approach for accurate time-series reconstruction of low-frequency PIV data [30, 31]. Giannopoulos and Aider [32] used neural network and time-resolved PIV data and successfully predicted temporal POD coefficients of the flow using visual sensors from local velocity measurements. They found a shallow neural network as a sufficient tool to obtain reasonable accuracy with low computational time. Liu et al. [33] utilized two deep learning models on DNS data of channel flow to develop spatial resolution enhancement. They obtained reproduction accuracy in capturing the anisotropic characteristics in the wall regions of flow field. The data resolution enhancement was also achieved after applying generative adversarial network to the PIV data of the complicated wake flow behind two side-by-side cylinders [34].

(4) *Reproduction of velocity data at the gappy regions by ML models*: Implementation of neural network on three artificially-generated flow data was used to filter the inherent noise and reconstruct the flow field at the gappy regions [35]. There was agreement between the predicted output and the theoretical target data for training a network based on either exact or noisy data. Utilization of neural network with radial basis function (RBF) was also reported as a robust interpolating method for reconstructing noisy data from PIV of a swirling flow [36]. The RBF interpolation scheme was also utilized for particle tracking velocimetry data and its outcome was superior to those of standard Gaussian weighting and Taylor expansion interpolation methods [37]. Extreme learning machine auto-encoders was another methodology applied on sparse flow data and found to be as a competitive reconstruction approach compared with the POD-based methods [38].

The above algorithms commonly utilize the spatial coordinate as the input feature of the ML regressors without implementation of flow-based quantities. Substantial capability of ML algorithms in reproduction of sparse flow fields, shows the potential for moving from a data-based approach towards a physics-based approach. The present study aims to use these potentials for proposing a novel approach in which flow-based features with practical fluid mechanics characteristics are considered in addition to the conventional coordinate-based features. The objective is reconstruction of the velocity vectors in any instantaneous PIV-measured turbomachinery flow field at the gappy regions in which spurious data are detected and removed. The path is to identify features and metrics that are native to fluid mechanics problems and facilitate ML algorithms. Multi-layer perceptron (MLP) and support vector regression are two ML models adopted due to their suitable performance in handling highly non-linear flow attributes in the complicated turbomachinery field. In addition to conventional evaluation of ML models based on the velocity data, a metric is proposed that relies on mass conservation principle as an important requisite for a physical reconstructed field.

The article is organized as follows. Section 2 introduces the overall procedure, including PIV data acquisition, data validation procedure, ML pre-processing steps, training of data by SVR and MLP, and hyper-parameter optimization. Model evaluation metrics are introduced in section 3. Results and discussion follow in section 4 presenting various aspects of flow field reconstruction and validation.

**2. PIV data acquisition, process and reconstruction strategy**

There are eight tasks to acquire initial experimental data as raw PIV images, process them into validated velocity vectors and reconstruct them to predict the missing data (as depicted in Figure 1). In the following, these steps are discussed in more details.

*2.1. Experimental setup, data acquisition and raw-image processing*

Stereoscopic particle image velocimetry (SPIV) is utilized to measure three instantaneous velocity components at various points of a planar Cartesian grid. The data is acquired from a constructed forward-blade centrifugal fan where images are recorded at an encoded position of the rotor. The experimental setup and instrumentation include the following components (as illustrated in Figure 1):

- A 43-blade centrifugal fan constructed from galvanized plate (Table 1)
- A test setup and an outlet duct compatible with ISO 5801 [39]
- A volute made of transparent Plexiglas to provide optical access to the test section
- A double-cavity Quantel Brilliant Nd-YAG laser equipped with an optical guide system that delivers the laser sheet to the test section
- Two FlowSense® 1600×1186 pixel double-frame CCD cameras
- SAFEX F2010 plus fog generator
- Dantec FlowMap® system for synchronization of laser, camera and encoder actions

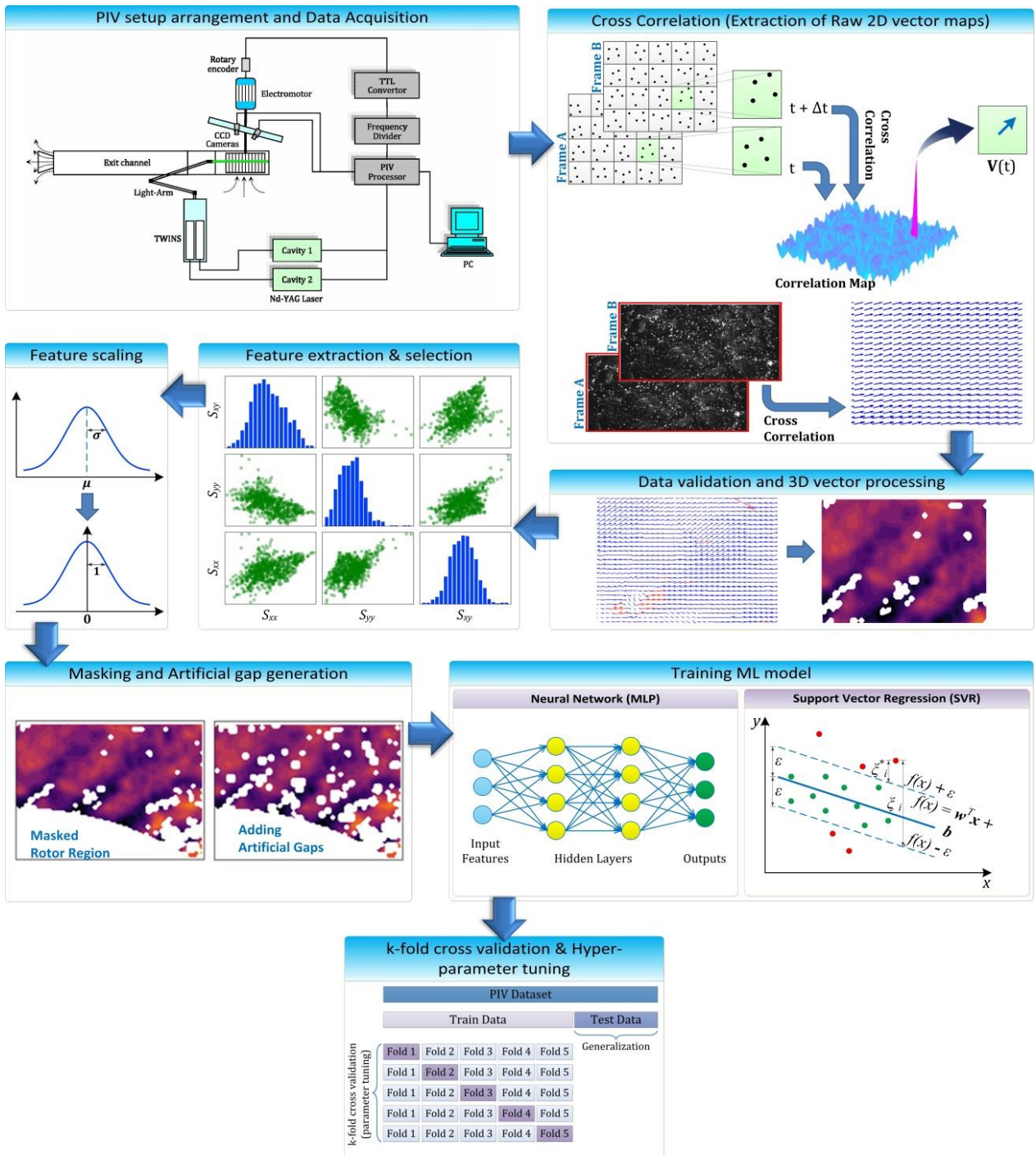

Figure 1. The flow diagram from data acquisition to reconstructed velocity map including eight steps: 1) PIV setup and image acquisition; 2) implementation of multi-stage adaptive cross-correlation; 3) data validation, calibration and 3D vector map calculation; 4) feature extraction and selection; 5) feature scaling; 6) masking and adding random artificial gaps to the flow field; 7) Training the SVR and MLP models; 8) Applying k-fold cross validation and hyper-parameter optimization.

Velocity measurement is performed at a plane perpendicular to the rotor axis at the upper region of the rotor with $z/B = 0.6$ where $z$ is the axial coordinate, measured from the volute inlet, and $B$ is the volute width (Figure 2(a)). Previous experimental studies on this centrifugal turbomachine indicate that the selected region involves the most intensive mutual interactions between jet-wake-volute flow structures [40, 41]. This complicated flow field would be a desirable candidate to evaluate performance of ML algorithms in correct reconstruction of the field at target regions. The rectangular measurement area is divided to four overlapping regions, termed as field of view (FOV), in order to both measure a fairly extensive region and resolve sufficiently detailed flow information (Figure 2(b)). The size of each FOV is $100 \times 74 mm^2$ with velocity vector spacing ($\delta$) in both $x$ and $y$ directions to be $1 mm$. The coordinate origin for each FOV is located at its lower left corner.

FlowManager® software [42] is used to extract the velocity components from two pairs of double-frame images. Multi-stage adaptive cross-correlation is utilized to obtain 2D vector map for each separate image. The acquired data consists of 2000 instantaneous snapshots for each FOV.

### 2.2. PIV data validation and 3D vector processing

The PIV vector maps after adaptive cross-correlation may contain some unreasonable and incorrect vectors, called "spurious vectors" or "outliers". They could appear scattered over an extended area or clustered at particular locations of flow field. The reasons for existing spurious vectors could be inhomogeneous dispersion of seeding particles, low intensity of laser illumination, optical noise and poor quality of a captured image. Prior to any velocity field analysis, it is important to detect and substitute spurious vectors, either scattered or clustered. Data validation procedure, including velocity-range validation and peak-height validation are employed to eliminate spurious velocity vectors [43].

Two concurrent 2D vector maps are then combined in order to achieve a united planar vector map including three velocity components for each instantaneous snapshot. The resulted vector map includes regions with missed data from the validation procedure that could be imputed by machine learning approach.

Table 1. Geometrical characteristics of the centrifugal fan [44].

| Region | Parameter | Value | Parameter | Value |
|---|---|---|---|---|
| **Rotor** | Inner diameter, $D_1$ | *285mm* | Outer diameter, $D_2$ | *350mm* |
| | Number of blades, $N$ | *43* | Rotor width, $b$ | *165mm* |
| | Blade entrance angle, $\beta_1$ | *90°* | Blade exit angle, $\beta_2$ | *155°* |
| **Volute** | Volute width, $B$ | *200mm* | Outlet height, $h$ | *200mm* |
| | Scroll spread angle, $\alpha$ | *5°* | Cut-off angle, $\gamma$ | *60°* |
| | Cut-off radius, $r_0$ | *210mm* | Rotor cut-off gap, | *35mm* |
| **Inlet** | Inner diameter, $d_1$ | *295mm* | Outer diameter, $d_2$ | *305mm* |
| | Curvature radius, $r_I$ | *15mm* | | |

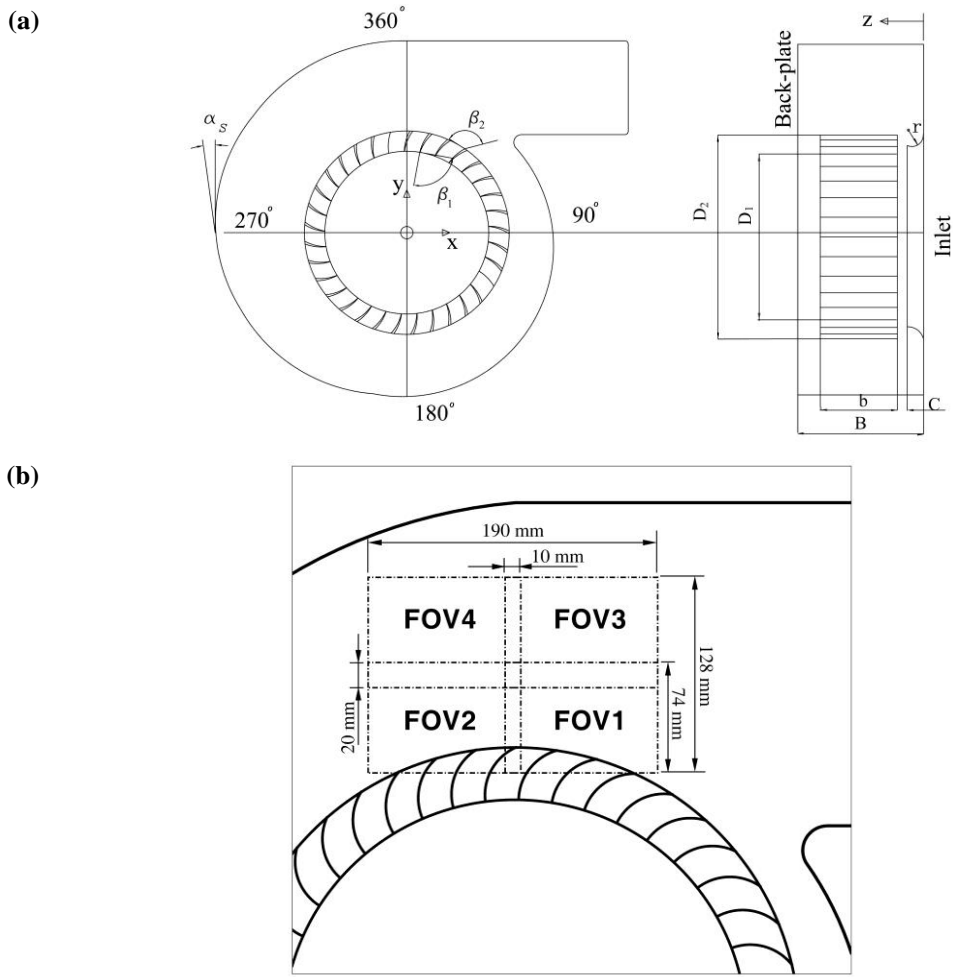

Figure 2. (a) Geometrical characteristics of the centrifugal fan; (c) Location and dimension of the four fields of view at the rotor exit region [44].

## *2.3. Flow-based feature extraction and scaling*

Data reconstruction in this paper pertains to different instantaneous snapshots, each contains three velocity components in the Cartesian grid of 98×73 with some missing data, either scattered or clustered. Each velocity component is the target of ML model for reconstruction, and appropriate input features should be provided for a robust training and prediction.

Because of the dependence of flow characteristics to geometry and position, the most basic input feature would be the spatial coordinate. It has been the conventional feature, utilized in the literature for reconstruction purpose [35-37]. The planar (*x* and *y*) coordinate of each velocity sample based on the regular PIV grid is consequently selected as the first category of input features, which is termed as 'type A' feature hereafter.

Implementation of suitable flow-based features with fluid mechanics significance would enhance the velocity reconstruction performance due to their inherent correlations with the flow field. The present study proposes two types of features in this regard: 'type B' are features characterizing by the measured velocity at the neighbouring positions and 'type C' are features relying on spatial gradient of velocity field at the target coordinate. Before describing these feature types, it is important to notice that any extracted feature should be non-zero in order to be used in a ML regressor. However, for clustered gappy data, existence of several null data in the neighbouring regions is quite possible. In order to impute such null features, it is proposed to apply filtering operation on both of the above-mentioned flow-based features over a customized rectangular domain.

'Type B' features are calculated after filtering the original PIV velocity field with different filter scales. In this way, the velocity data at the proximity of each target point contributes to the calculated feature. It takes advantage of similarity between neighbouring velocity vector, particularly where the target point with null data is surrounded by non-zero velocity vectors. The general filtering operation on each instantaneous velocity component over a spatial domain $D$ is given as [45]:

$$\overline{u_i}(x_1, x_2) = \iint_D G(x_1 - x_1', x_2 - x_2') u_i(x_1', x_2') dx_1' dx_2' \quad (1)$$

where the overbar denotes the filtering operation, $x_1$ and $x_2$ are spatial coordinates in x and y directions and $G$ is the filter kernel. $i$ indicates the velocity components which are 1, 2 and 3 for *x*, *y* and *z* directions, respectively. The top-hat (box) and Gaussian kernels have been frequently utilized in fluid dynamics scope [46] which are selected in our study as well. Because of discrete values of PIV velocity, the continuous filtering is not applicable and a discontinuous form should be utilized. The discrete kernels applied for the top-hat and Gaussian filters are [45]:

$$G_{top-hat}(x_1, x_2) = \begin{cases} C_1 & \text{if } |x_1| < \Delta_x \text{ and } |x_2| < \Delta_y \\ 0 & \text{otherwise} \end{cases} \quad (2)$$

$$G_{Gauss}(x_1, x_2) = C_2 \exp\left[-\frac{6(x_1^2 + x_2^2)}{\Delta_x \Delta_y}\right] \quad (3)$$

$\Delta_x$ and $\Delta_y$ are the filter scales in x and y directions, respectively. The coefficients $C_1$ and $C_2$ are normalization factors which ensure that the integral of the filter kernel over the filtering region equals unity (in a discrete manner). The null velocity at gappy locations should be excluded from filtering operation. Consequently, $C_1$ and $C_2$ are dependent to the count and position of non-zero velocities.

The 'type C' features are defined to be the filtered deformation-based quantities (filtered strain rate tensor and filtered vorticity vector) as frequently-used attributes in fluid mechanics. The filtered strain rate tensor ($\mathbf{S}$) as the symmetric part of filtered deformation tensor is defined as [47]:

$$\overline{\mathbf{S}} = \frac{1}{2}\left(\nabla \overline{\mathbf{V}} + \nabla \overline{\mathbf{V}}^T\right) \quad (4)$$

where $\overline{\mathbf{V}}$ is the filtered velocity vector and the superscript $T$ notifies the matrix transpose operation. Furthermore, the filtered vorticity vector ($\overline{\boldsymbol{\omega}}$) as the anti-symmetric part of filtered deformation tensor is calculated as curl of the filtered velocity vector:

$$\overline{\boldsymbol{\omega}} = \nabla \times \overline{\mathbf{V}} \quad (5)$$

Because of planar nature of PIV measurements, the in-plane components of the filtered strain rate tensor (Equation 4) and the out-of-plane component of the vorticity vector (Equation 5) are calculable.

Since the range of values for three proposed feature types are different, prior to importing them to the ML regressors, it is necessary to scale the features by standard methods. Four conventional approaches are applied in the present study for this purpose: 1) 'normalizer' method that scales each individual sample to the unit norm; 2) 'min-max scaler' that transforms each feature by scaling it to a customized range; 3) 'standard scaler' that standardizes each feature vector to a distribution with zero mean and unit variance; 4) 'robust scaler' that scales the feature vectors based on customized quantile range and is robust to outliers.

*2.4. Artificial gap generation*

In order to evaluate the ML model accuracy, it is necessary to add some artificial gaps to provide unseen data for the trained model. The gap percentage, $GP$, is defined as the ratio of the number of gappy data in the active (non-masked) region of each snapshot to the overall number of points in that region. The initial value of gap percentage depends on the extent of spurious vectors. A gappiness threshold (in percent), $GT$, is the target situation in which the extent of gappy data (including the artificial gaps) reaches the desired quantity. The higher the assigned gappiness threshold, the more challenging accurate prediction of velocity data in the gappy regions is. Utilizing a gap generator, we add numerous rectangular masks with random dimensions in specified range and apply them in randomly adopted locations through the FOV, until reaching to the customized gappiness threshold ($GP \geq GT$). Considering the space pertaining to artificially generated gappy data as $\mathbf{S}_G$, the masked velocity components ($u_i^{\text{mask}}, i=1,2,3$) are given as:

$$u_i^{\text{mask}} = \begin{cases} 0 & \text{if } \mathbf{R} \in \mathbf{S}_G \\ u_i^{\text{orig}} & \text{otherwise} \end{cases} \qquad (6)$$

where $u_i^{\text{orig}}, i=1,2,3$ is the original velocity components (including the detected spurious velocity vectors as null data) and $\mathbf{R}$ is the two-dimensional spatial vector (coordinate) with respect to the FOV origin. We utilize a portion of flow data from the space $\mathbf{S}_G$ with given velocity components as 'test data' to evaluate the ML model accuracy.

After isolating the test data, the remainder is randomly split between two categories: the training data and the validation data. The training data is utilized to calculate the weight coefficients in each ML model while the validation portion is used to evaluate the model accuracy and tuning the model hyper-parameters.

*2.5. Training the machine learning models*

Two conventional machine learning approaches are utilized to model and reconstruct the velocity field:

(1) **Support vector machine**: Support vector machine, originated by Vapnik et al. [48], is one of the most popular algorithms in supervised learning, either for regression or classification. SVM transforms the input data into a high dimensional feature space through nonlinear mapping and then performs a linear regression in that feature space. Using the so-called kernel trick approach to capture non-linear characteristics of the feature space, SVM provides a robust model that is suitable for prediction of complex data sets with small to medium size [49]. The SVM regression model (SVR) can be obtained from training a quadratic optimization problem for specified hyper-parameters, namely $C$ and $\varepsilon$ ($C$ is a positive coefficient for the degree of penalized loss for the customized radius of the insensitive tube $\varepsilon$). The selected kernel in the present study is the frequently-used Gaussian (radial basis function, RBF) kernel with $\gamma$ as a hyper-parameter characterizing the inverse of the influence radius of support vectors.

(2) **Multi-layer perceptron**: Feed-forward multi-layer perceptron (MLP) neural network is the second approach applied to reconstruct the PIV velocity field. The input layer consists of the learning features and the output layer includes the velocity components. The rectifier linear unit (ReLU) function is implemented as the activation function because it avoids and rectifies the vanishing gradient problem in the gradient descent optimization of weights. Moreover, the ReLU is less computationally expensive than other well-known activation functions such as sigmoid and hyperbolic tangent. Back-propagation algorithm is applied to minimize the error between the predicted output of the MLP model and the desired output in order to calculate the synaptic weights [49].

There are several merits for selecting these two models as estimators. Implementation of numerous layers in conjunction with applying non-linear activation functions in MLP and applying the SVR kernel with non-linear characteristics introduce them as two powerful tools in estimation of the complicated fluid flow phenomena for moderate amount of data [22]. Flexibility of MLP in learning process regardless of network structure and robustness of SVR against overfitting, particularly in higher dimensions of feature space, and its reasonable computational cost are the other supportive reasons for such selection.

## *2.6. Optimization of hyper-parameters via k-fold cross-validation*

Optimal values of model hyper-parameters should avoid overfitting and lead to suitable generalization performance for the model. Grid search and random search are two frequently used schemes to this end. The random search approach was proved empirically and theoretically more efficient for hyper-parameter optimization than the grid search scheme [50] and is consequently applied in the present study.

Based on the random search scheme, a wide range of hyper-parameters are considered and a series of ML models are trained. A k-fold cross validation procedure is employed to evaluate the model performance and in order to avoid occurrence of overfitting. After splitting training data set into $k$ subsets of equal size, each subset is used for the model evaluation by feeding the remaining $k-1$ subsets as the input data to the model for training purpose. This is iterated until each of $k$ subsets was used for validation. The model is then evaluated based on a suitable error metric for each of $k$ validation subsets and the average of calculated errors is considered as the representative error for the hyper-parameter combination used. The minimum error among all different combinations is selected after comparing the averaged error metric calculated for various randomly adopted hyper-parameters.

## 3. Model evaluation and error metrics

Several error metrics could be applied to evaluate the model accuracy either in the validation or test stages. Since the PIV data contains numerous outliers, the median absolute error ($MedAE$) of each velocity component is the first metric selected for model evaluation because it is less sensitive to outliers in comparison with conventional metrics such as mean absolute error and mean squared error. Furthermore, because of different scales of the three velocity components (as the regressor target), the median of all normalized absolute errors between target and predicted values is regarded as *MedNAE*, and is defined as follows:

$$MedNAE = median\left(\frac{|y_1 - \bar{y}_1|}{|y_1|}, \frac{|y_2 - \bar{y}_2|}{|y_2|}, \ldots, \frac{|y_{N_s} - \bar{y}_{N_s}|}{|y_{N_s}|}\right) \tag{7}$$

where $y$ and $\bar{y}$ denote the target and predicted values of the velocity components, respectively and $N_s$ is the number of data samples. The closer to zero is this error metric, the more accurate the predictor performs.

The second frequently-used metric is the coefficient of determination, also termed as $R^2$-score. It is expressed as:

$$R^2 = 1 - \sum_{i=1}^{N_s}(y_i - \bar{y}_i)^2 \Big/ \sum_{i=1}^{N_s}(y_i - \langle y \rangle)^2 \tag{8}$$

where $\langle y \rangle$ is the average of target values, $\langle y \rangle = \left(\sum_{i=1}^{N_s} y_i\right) \Big/ N_s$. This error metric indicates how good the model fits and how good it expresses the amount of variability in the target quantity that is accounted for by the predictor variables [51]. Best possible score of $R_2$ is unity, while the value of zero pertains to the case in which the model always predicts the same value, regardless of the input feature values (an uncorrelated model).

In addition to the above mentioned velocity-based metrics, as a third metric we propose to evaluate the mass conservation for the original and reconstructed velocity fields as a fundamental fluid mechanics requirement. For an incompressible flow with continuous velocity field, the mass conservation equation is given as follows:

$$\nabla \cdot \mathbf{V} = 0 \tag{9}$$

In the case of planar PIV data, the in-plane velocity gradients are calculable by finite difference approach and based on the neighboring velocity components. Accordingly, the expanded form of Equation 9 could be written as follows, in which the right-hand-side terms are directly calculable from the planar velocity data:

$$\left(\frac{\partial u_3}{\partial x_3}\right)^{(i,j)} = -\left(\frac{u_1^{(i+1,j)} - u_1^{(i-1,j)}}{2\delta_x} + \frac{u_2^{(i,j+1)} - u_2^{(i,j-1)}}{2\delta_y}\right) \tag{10}$$

The subscripts 1, 2 and 3 in Equation 10 denote the *x*, *y* and *z* coordinates, respectively. The superscripts *i* and *j* notify the *x* and *y* positions of the target point in the discrete PIV map. The left-hand-side term in Equation 10 could be considered as a metric that evaluates the out-of-plane mass flux in a conservative way. This is termed as mass conservation factor ($MC$) hereafter. Since the final objective is to reconstruct a planar velocity field in a consistent manner to the real two-dimensional PIV field, a higher ML accuracy occurs when $MC$, calculated from the reconstructed velocity field is as close as possible to that of the original PIV data. After calculation of $MC$ for all test points of different FOVs and snapshots, model accuracies in terms of mass conservation is evaluated using conventional statistical metrics, namely median absolute error and $R^2$-score.

The scores calculated for each of the four FOVs, are averaged for all snapshots in order to stabilize their trends and extract the general effect of various parameters in performance of each ML model. Our examinations indicated that using 100 snapshots is sufficient to obtain a stable and converging trend for each FOV.

## 4. Results and Discussion

### *4.1. Contribution of flow-based input features*

The velocity components obtained by the ML models are evaluated for three different combinations of input features:

(1) Combination A: The x- and y-coordinate of each velocity component, i.e. 'type A' features (based on definitions of section 2.3)
(2) Combination B: The velocity vectors coordinate plus the filtered velocity components, i.e. 'type A + type B' features
(3) Combination C: The velocity components coordinate plus the filtered velocity plus the filtered vorticity/strain components, i.e. 'type A + type B + type C' features.

Table 2 examines the effect of these three feature combinations on the accuracy of velocity component reconstructions by SVR and MLP models in term of MedNAE and $R_2$-Scores. Our main concern is to investigate whether employing flow-based features in addition to the coordinate-based features enhances the prediction accuracy. Prior to hyper-parameter optimization (section 4.6), the following configuration of hyper-parameters is considered for two implemented ML models: $C = 50$, $\gamma = 1$ and $\varepsilon = 0.01$ for the SVR, and 32 neurons in the first and 64 neurons in the second layers of the MLP with two hidden layers. Based on Table 2, different trends are observed for SVR and MLP outcomes, particularly when combinations A and B are compared. More specifically, although adding the filtered velocity components as predictor features is quite beneficial in growth of MLP model accuracy, such configuration does not enhance the accuracy of SVR model. Low accuracy of SVR could be addressed by strong sensitivity of this ML model to the features selected and their scales [52]. It will be shown that utilization of appropriate scaling scheme circumvents such problem. Regarding the MLP model, negative values of $R_2$-Score in combination A indicates that the coordinates as the input feature are not enough for this model to correctly predict the velocity field. However, adding the filtered velocity components as predictive features increases the $R_2$-Score to above 0.75 for all velocity components and decreases the MedNAE by an order of magnitude. This indicates the important role of flow-based quantities as the predictors.

Table 2. Contribution of various predictive features on the performance of SVR and MLP models in reconstruction of separate velocity components.

| ML model | Velocity component | Active Features | | | | | |
|---|---|---|---|---|---|---|---|
| | | Combination A | | Combination B | | Combination C | |
| | | MedNAE | $R_2$-Score | MedNAE | $R_2$-Score | MedNAE | $R_2$-Score |
| SVR | 1 | 0.053 | 0.413 | 0.057 | 0.370 | 0.077 | -0.019 |
| | 2 | 0.180 | 0.410 | 0.182 | 0.377 | 0.245 | -0.010 |
| | 3 | 0.512 | 0.399 | 0.532 | 0.329 | 0.669 | -0.008 |
| MLP | 1 | 0.122 | -2.790 | **0.015** | **0.910** | 0.030 | -2.180 |
| | 2 | 0.251 | -0.131 | **0.066** | **0.908** | 0.114 | 0.417 |
| | 3 | 0.762 | -0.018 | **0.366** | **0.755** | 0.645 | -0.048 |

Table 2 also shows the sharp reduction of both model accuracies after inclusion of velocity deformation based features (the filtered vorticity/strain components). The orders of magnitude of vorticity and strain components are considerably different from those of velocity components and coordinate index. According to this issue and the results of combination C, it is necessary to scale the input features in order to make them more consistent and effective in achieving a better performance.

Table 3 evaluates the contribution of flow-based features to the accuracy of SVR and MLP reconstructions after applying various scalers on the data, namely the normalizer, min-max, standard and robust scalers. The results are compared for three combinations of input features: A, B and C. A comparison is also made with the predictions achieved without scaling procedure. A general enhancement of error metrics is evident as the result of feature scaling prior to the training procedure, particularly in combination C which includes features with a diverse range of values and patterns.

Comparison of results obtained for three different feature combinations in presence of scaling in terms of both $R_2$-Score and MedNAE, indicates that combinations including flow-based features (combinations B and C) result in higher accuracy compared to the coordinate-based features (combination A). Among these flow-based configurations, the combination B which does not include velocity deformation based features (vorticity and strain components) considerably outperforms the combination C. A possible reason would be the lower fluctuations of 'type B' features (which are calculated after filtering the velocity field) in comparison with 'type C' features which are derivative-based quantities.

Table 3. Contribution of different scaling methods on the performance of SVR and MLP models in prediction of separate velocity components using various predictive feature configurations.

| Active Features | ML model | Vel. Comp. | None | | Normalizer | | Min-Max Scalar | | Standard Scalar | | Robust Scalar | |
|---|---|---|---|---|---|---|---|---|---|---|---|---|
| | | | MedNAE | $R_2$-Score | MedNAE | $R_2$-Score | MedNAE | $R_2$-Score | MedNAE | $R_2$-Score | MedNAE | $R_2$-Score |
| Combination A | SVR | 1 | 0.053 | 0.413 | 0.047 | 0.454 | 0.039 | 0.643 | 0.022 | **0.846** | 0.032 | 0.695 |
| | | 2 | 0.180 | 0.410 | 0.163 | 0.380 | 0.144 | 0.666 | 0.075 | **0.878** | 0.118 | 0.762 |
| | | 3 | 0.512 | 0.399 | 0.658 | 0.150 | 0.576 | 0.311 | 0.458 | **0.572** | 0.536 | 0.416 |
| | MLP | 1 | 0.122 | -2.790 | 0.050 | 0.465 | 0.043 | 0.609 | 0.028 | **0.816** | 0.031 | 0.735 |
| | | 2 | 0.251 | -0.131 | 0.158 | 0.410 | 0.127 | 0.704 | 0.072 | **0.899** | 0.074 | 0.876 |
| | | 3 | 0.762 | -0.018 | 0.665 | 0.162 | 0.570 | 0.376 | 0.452 | **0.619** | 0.495 | 0.562 |
| Combination B | SVR | 1 | 0.057 | 0.370 | 0.035 | 0.701 | 0.011 | 0.851 | 0.010 | **0.946** | 0.011 | 0.942 |
| | | 2 | 0.182 | 0.377 | 0.094 | 0.805 | 0.035 | 0.945 | 0.031 | **0.971** | 0.034 | 0.965 |
| | | 3 | 0.532 | 0.329 | 0.321 | 0.749 | 0.227 | 0.834 | 0.227 | **0.847** | 0.228 | 0.817 |
| | MLP | 1 | 0.015 | 0.910 | 0.035 | 0.737 | 0.011 | 0.938 | 0.011 | **0.950** | 0.011 | 0.924 |
| | | 2 | 0.066 | 0.908 | 0.094 | 0.826 | 0.034 | 0.971 | 0.034 | **0.975** | 0.034 | 0.974 |
| | | 3 | 0.366 | 0.755 | 0.322 | 0.743 | 0.234 | **0.862** | 0.244 | 0.841 | 0.239 | 0.835 |
| Combination C | SVR | 1 | 0.077 | -0.019 | 0.049 | 0.468 | 0.010 | **0.902** | 0.018 | 0.725 | 0.015 | 0.762 |
| | | 2 | 0.245 | -0.010 | 0.150 | 0.551 | 0.032 | **0.957** | 0.060 | 0.711 | 0.050 | 0.773 |
| | | 3 | 0.669 | -0.008 | 0.480 | 0.466 | 0.225 | **0.847** | 0.340 | 0.591 | 0.310 | 0.654 |
| | MLP | 1 | 0.030 | -2.180 | 0.044 | 0.568 | 0.013 | 0.931 | 0.011 | 0.928 | 0.011 | **0.942** |
| | | 2 | 0.114 | 0.417 | 0.120 | 0.719 | 0.043 | 0.957 | 0.035 | 0.970 | 0.036 | **0.959** |
| | | 3 | 0.645 | -0.048 | 0.424 | 0.583 | 0.245 | **0.860** | 0.256 | 0.830 | 0.255 | 0.834 |

Based on the results presented in Table 3, the standard, robust and min-max scalers cause significantly better scores compared with those of the normalizer scaler. Considering combination C as the active feature configuration, the min-max scaler leads to more accurate reconstruction of velocity field. This is due to very different scales of vorticity/strain compared to types A and B features. The min-max scaler has a complete control to the output scaling range and consequently is more successful for those features in comparison with the standard and robust scalers that do scaling based on the distribution attributes of each feature. However, comparison of the best scores in Table 3 (with bold figures) confirms that the combination B of features and utilization of standard scaler (and robust scaler by a little difference) lead to the most accurate reconstruction. High performance of standard scaler method is expected to be due to normal distribution of velocity data, while capability of robust scaler in diminishing the contribution of outlier data could be the main reason for successful prediction by this scaling method. In the following sections of the article, feature combination B processed by standard scaler are utilized.

*4.2. Filtering*

Two box filters, one with uniform and the other with Gaussian weights, are implemented in 'configuration B' features and their effect as well as the contribution of filter length scale is presented in Table 4 in terms of $R_2$-Score. Each integer in the curly brackets indicates the filter size which is an integer multiplier to the PIV grid spacing in both x and y directions. For instance, {3, 5} specifies two square filter boxes by dimensions $3\delta \times 3\delta$ and $5\delta \times 5\delta$. There are random artificial gappy squares underneath each filter box and throughout each FOV, as described in section 2.4, which are applied with a range of dimensions from 2×2 to 4×4.

The $R_2$-score in the case of {3, 5} is significantly lower than the predictions made by the other filter scales. The minimum filter scale in the case of {3, 5} is 3×3 which is smaller than the size of some gappy boxes and covers just the null values in such positions. In order to select a primary configuration from Table 4 for the following sections, the absolute difference between each score with the maximum score in the same row is presented in Table 5. Zero values indicate the maximum score in the corresponding row. The items with the difference smaller than 0.005 (0.5%) are specified in bold. Although most items with maximum score are in the {7} configuration, the overall accuracy of {7, 9} seems to be slightly better (particularly regarding the third velocity component predicted based on the Gaussian kernel) which is consequently utilized in the subsequent sections. The lower overall accuracy in the case of {5, 7, 9} (excluding the {3, 5} configuration) indicates that utilizing more filter-based features does not necessarily enhance the model performance.

Table 4; Effect of different filter kernels and scales on the $R_2$-Score of SVR and MLP models in prediction of separate velocity components.

| Filter kernel | ML model | Vel. Comp. | Filter length scales | | | | | |
|---|---|---|---|---|---|---|---|---|
| | | | {3, 5} | {5, 7} | {7, 9} | {9, 11} | {7} | {5, 7, 9} |
| Uniform | SVR | 1 | 0.7415 | 0.9460 | **0.9554** | 0.9516 | **0.9595** | 0.9451 |
| | | 2 | 0.8023 | 0.9693 | **0.9783** | 0.9773 | **0.9799** | 0.9719 |
| | | 3 | 0.5666 | **0.8431** | **0.8473** | 0.7874 | 0.8307 | 0.8198 |
| | MLP | 1 | -0.9582 | 0.9391 | **0.9489** | 0.9383 | **0.9527** | 0.9366 |
| | | 2 | 0.5881 | **0.9718** | **0.9717** | 0.9661 | **0.9725** | **0.9684** |
| | | 3 | 0.5063 | **0.8390** | 0.8126 | 0.7713 | 0.8195 | **0.8414** |
| Gaussian | SVR | 1 | 0.7456 | 0.9458 | **0.9571** | 0.9584 | **0.9579** | 0.9465 |
| | | 2 | 0.8050 | 0.9710 | **0.9796** | 0.9805 | **0.9819** | 0.9724 |
| | | 3 | 0.6093 | **0.8474** | **0.8513** | 0.8368 | 0.8339 | 0.8385 |
| | MLP | 1 | -1.2556 | **0.9497** | 0.9503 | 0.9452 | **0.9509** | 0.9481 |
| | | 2 | 0.5764 | **0.9751** | 0.9747 | 0.9699 | **0.9747** | 0.9729 |
| | | 3 | 0.5506 | **0.8406** | **0.8398** | 0.7969 | 0.8096 | 0.8270 |

Table 5. The absolute difference between each score in Table 4 with the maximum score existed in the same row.

| Filter kernel | ML model | Vel. Comp. | Filter length scales | | | | | |
|---|---|---|---|---|---|---|---|---|
| | | | {3, 5} | {5, 7} | {7, 9} | {9, 11} | {7} | {5, 7, 9} |
| Uniform | SVR | 1 | 0.2181 | 0.0135 | **0.0041** | 0.0079 | **0** | 0.0145 |
| | | 2 | 0.1777 | 0.0106 | **0.0017** | **0.0026** | **0** | **0.0081** |
| | | 3 | 0.2807 | **0.0042** | **0** | 0.0600 | 0.0167 | 0.0276 |
| | MLP | 1 | 1.9109 | 0.0136 | **0.0039** | 0.0145 | **0** | 0.0161 |
| | | 2 | 0.3844 | **0.0007** | **0.0008** | 0.0064 | **0** | **0.0041** |
| | | 3 | 0.3351 | **0.0024** | 0.0287 | 0.0701 | 0.0218 | **0** |
| Gaussian | SVR | 1 | 0.2128 | 0.0126 | **0.0013** | **0** | **0.0005** | 0.0119 |
| | | 2 | 0.1770 | 0.0109 | **0.0024** | **0.0015** | **0** | 0.0095 |
| | | 3 | 0.2420 | **0.0039** | **0** | 0.0145 | 0.0174 | 0.0128 |
| | MLP | 1 | 2.2066 | **0.0012** | **0.0006** | 0.0058 | **0** | **0.0029** |
| | | 2 | 0.3987 | **0** | **0.0004** | 0.0052 | **0.0004** | **0.0022** |
| | | 3 | 0.2899 | **0** | **0.0008** | 0.0437 | 0.0309 | 0.0136 |

In terms of filter kernel shape, Table 4 shows that slightly better predictions are made by the Gaussian weights as compared with those of the uniform box filter. This is because of weight coefficient reduction in the Gaussian kernel away from the target point, while the uniform kernel considers similar weights for all in-box points.

### 4.3. Velocity field reconstruction

Figures 3 to 5 depict contour plots of the three instantaneous velocity components in the rotor exit region. Each of these three figures contains four parts: (a) the originally measured velocity field by SPIV; (b) a velocity map after artificial generation of numerous random gap boxes with dimensions ranging from 2×2 to 4×4 and 25% gappiness; (c) velocity field prediction by SVR; (d) the reconstructed velocity field by MLP.

All four FOVs are included in Figures 3 to 5 because of their different flow attributes. The rotor exit-flow, which is dominant in FOVs 1 and 2, is significantly cyclic due to jet-wake flow structures after the blades. Moving downstream towards FOVs 3 and 4, the high-momentum jets and low-momentum wakes have mixed and produced more uniform flow patterns in the volute. These patterns are particularly evident from the in-plane velocity components (Figures 3 and 4), while the out-of-plane velocity component has more complicated and fluctuating pattern due to three-dimensional effects in the rotor-exit flow of this type of turbomachine [40]. The diversity of characteristics mentioned for different velocity components at four FOVs provides an appropriate collection of data for ML model evaluations.

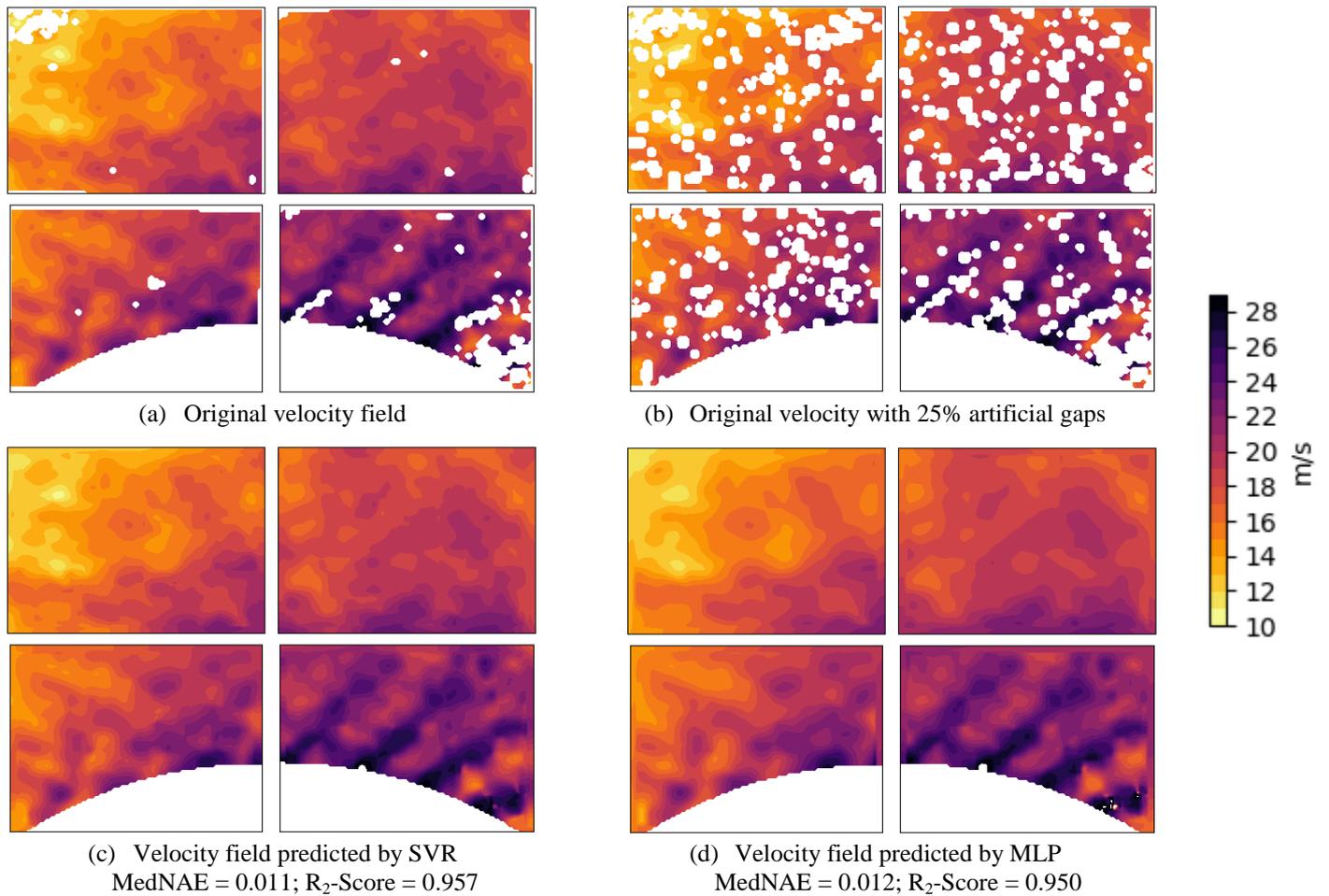

(a) Original velocity field

(b) Original velocity with 25% artificial gaps

(c) Velocity field predicted by SVR
MedNAE = 0.011; $R_2$-Score = 0.957

(d) Velocity field predicted by MLP
MedNAE = 0.012; $R_2$-Score = 0.950

Figure 3. The x-component velocity field in the four rotor exit FOVs

The MedNAE and $R_2$-Scores for the SVR and MLP predictions are also presented in parts (c) and (d) of Figures 3 to 5. In terms of these metrics, the SVR and MLP models have similar accuracies, slightly in favour of SVR. Comparison of identical iso-velocity contours in parts (c) and (d) of each figure with those of part (a) as the target indicates that MLP regressor generates more smooth velocity patterns, in comparison to the SVM model which reconstructs the original velocity patterns with more detail. From a machine learning viewpoint, precise reconstruction of the flow field patterns, particularly in cases with more fluctuative trends, could cause two drawbacks: 1) it makes the ML model more complicated and consequently rises the computational cost; 2) overfitting is more likely when a more complex model (namely SVM with more complicated kernel or MLP with deeper hidden layers and more number of neurons in each hidden layer) are utilized. However, regarding the fluid dynamics viewpoint, a more accurate reconstruction of the original wavy pattern is crucial, due to the sensitivity of velocity-derived quantities to reconstruction accuracy. Consequently if the overfitting problem is avoided, a more complicated and more accurate ML model is plausible. This issue is confirmed in section 4.5 by comparing the scores calculated in the training, cross-validation and testing stages and also in section 4.6.2 by evaluating the accuracy in terms of mass conservation factor.

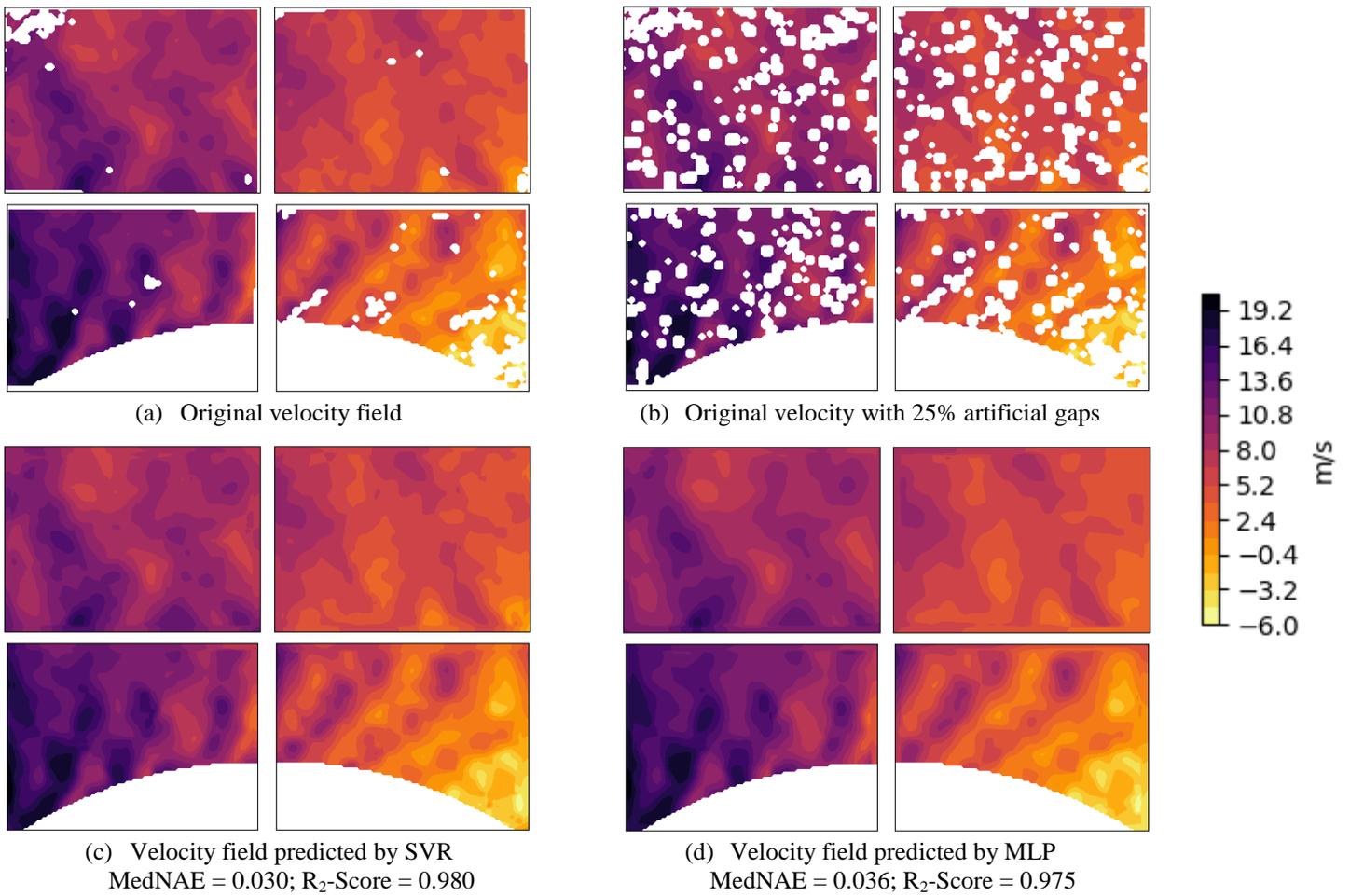

Figure 4. The y-component velocity field in the four rotor exit FOVs.

Possible size of reproducible gappy regions depends on the selected features and their availability in the target gappy locations. For combination B, with 7×7 and 9×9 filter dimensions, models are capable of reconstructing gappy regions as large as a 7×7 square box. According to Figures 3 to 5, even at challenging locations of FOVs 1 and 4 with large spots of gappy data, ML models are successful in full reconstruction of the velocity field. More specifically, it is noticeable from the prediction for the bottom right corner of FOV 1 in Figure 3 and the top left corner of FOV 4 in Figure 4 that the ML models have successfully reconstructed the jet-wake flow pattern in a consistent way, although most of data in those regions were originally missing.

Figure 6 illustrates reconstruction of three velocity components of an instantaneous snapshot of FOV 2 with sparse data and gappiness of 50% including random gap boxes with dimensions from 2×2 to 6×6. Considering the extensive regions with artificial null data, the overall capability of SVR and MLP models in predicting a similar pattern to that of the original velocity is still satisfactory, and slightly in favour of the SVR model. There are some local jumps in the predicted values, particularly in the MLP model, that correspond to locations with wider artificial gappiness. Utilization of broader filters could remove such discontinuities for a drop in overall model accuracy.

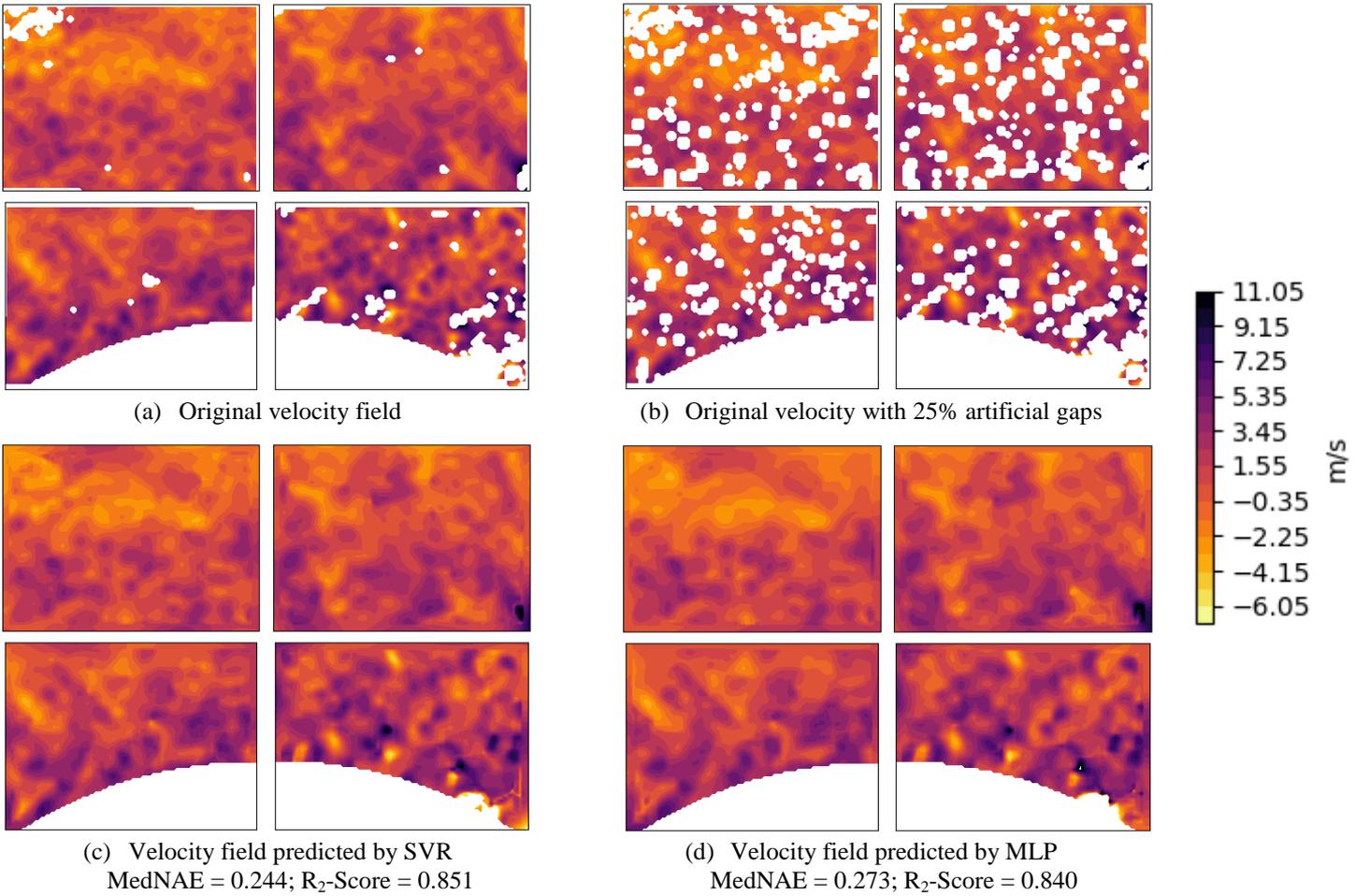

(a) Original velocity field

(b) Original velocity with 25% artificial gaps

(c) Velocity field predicted by SVR
MedNAE = 0.244; $R_2$-Score = 0.851

(d) Velocity field predicted by MLP
MedNAE = 0.273; $R_2$-Score = 0.840

Figure 5. The z-component velocity field in the four rotor exit FOVs:

In order to examine the velocity profile predicted by the ML models in more quantitative manner, Figure 7 demonstrates the output of Figure 6 evaluated at the horizontal dashed line shown in that figure (the line $y = 40\delta$, where $\delta$ is the PIV grid spacing and the coordinate origin is located at the lower left corner of the FOV). This figure compares SVR and MLP predictions of velocity (solid curves) with the original velocity as well as the artificially gap-augmented data (scattered symbols). The velocity at locations with originally missing data (due to PIV data validation) or locations selected as artificial gaps (for model evaluation) is zero and otherwise, it is the same as the original measured velocity. The reproduction of the true in-plane velocity profile in the gappy points is considerably more accurate than the reconstruction of the out-of-plane velocities. This is in agreement with the scores and trends presented previously in Figures 3 to 5 for lower level of gappiness percentage. The wavier pattern of the out-of-plane velocity component is the main reason for achieving a lower accuracy of reconstruction.

Despite alternating pattern of the y-component of velocity and large bundles of artificial missing data, namely between $x = 23\delta$ and $x = 35\delta$, both SVM and MLP have been quite successful in prediction of the velocity peak at around $x = 30\delta$ and reconstruction of the steep changes at its side regions. Furthermore, the discontinuity in the x-component velocity profile induced by gappiness of the original data between $x = 55\delta$ and $x = 57\delta$ is located at a high-gradient region of the velocity field. The predicted trend by the ML models at this interval is accurate. Both predictions show local peaks which indicate large scale turbulence which are not evident in point to point velocity values but were also repeated in iso-velocity contours.

### *4.4. Assessment of the gap effect*

Capability of SVR and MLP in reconstruction of two-dimensional instantaneous vector maps in FOV2 is investigated in Figure 8 for two extensively sparse fields with different levels of gappiness percentage, namely 50% and 70%. Such broadly sparse missing data are possible for PIV data of complex flow fields, when seeding concentration is non-homogeneous and insufficient in some clustered regions. The blue vectors in Figure 8 indicate the original measured velocity, while the red ones denote the reconstructed vectors by the ML models. Although growth of gappiness slightly increases the deviation between true and predicted vectors, reasonable coincidence of vectors, even for largest amount of gappiness, indicates good potential of both ML models in correct prediction of vector maps. However, more detailed comparison of the results indicates marginally better coincidence for the vectors obtained by SVR than those of MLP. This is also confirmed by the $R_2$-scores reported in Figure 8 for each vector plot calculated based on the planar velocity magnitude.

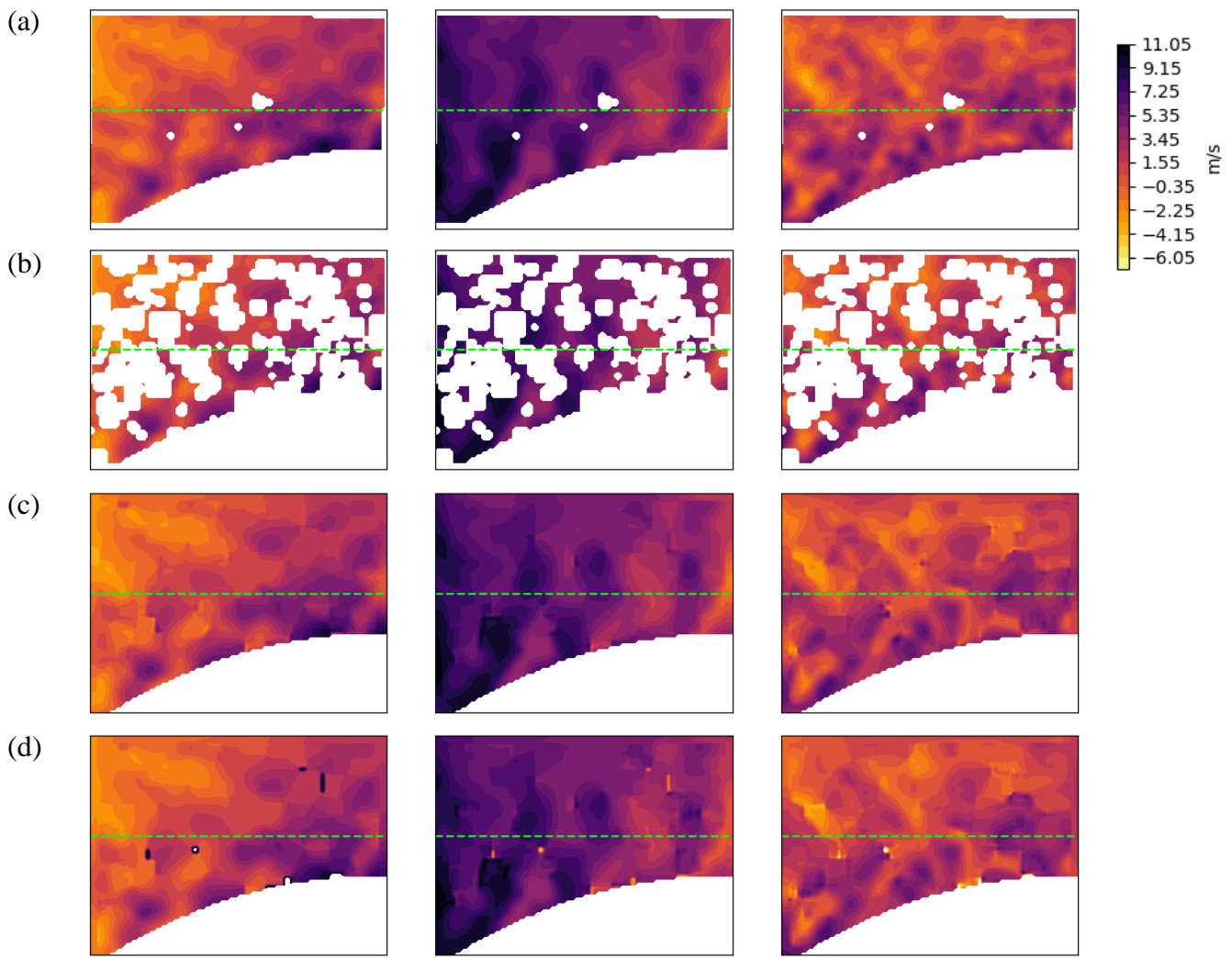

Figure 6. The x-, y- and z-velocity components (left, middle and right columns) of an instantaneous snapshot of FOV 2: (a) Original velocity measured by SPIV; (b) Original SPIV velocity field plus 50 percent of artificial gappiness applied with a range of dimensions from 2×2 to 6×6; (c) the predicted velocity field by SVR; (d) the predicted velocity field by MLP.

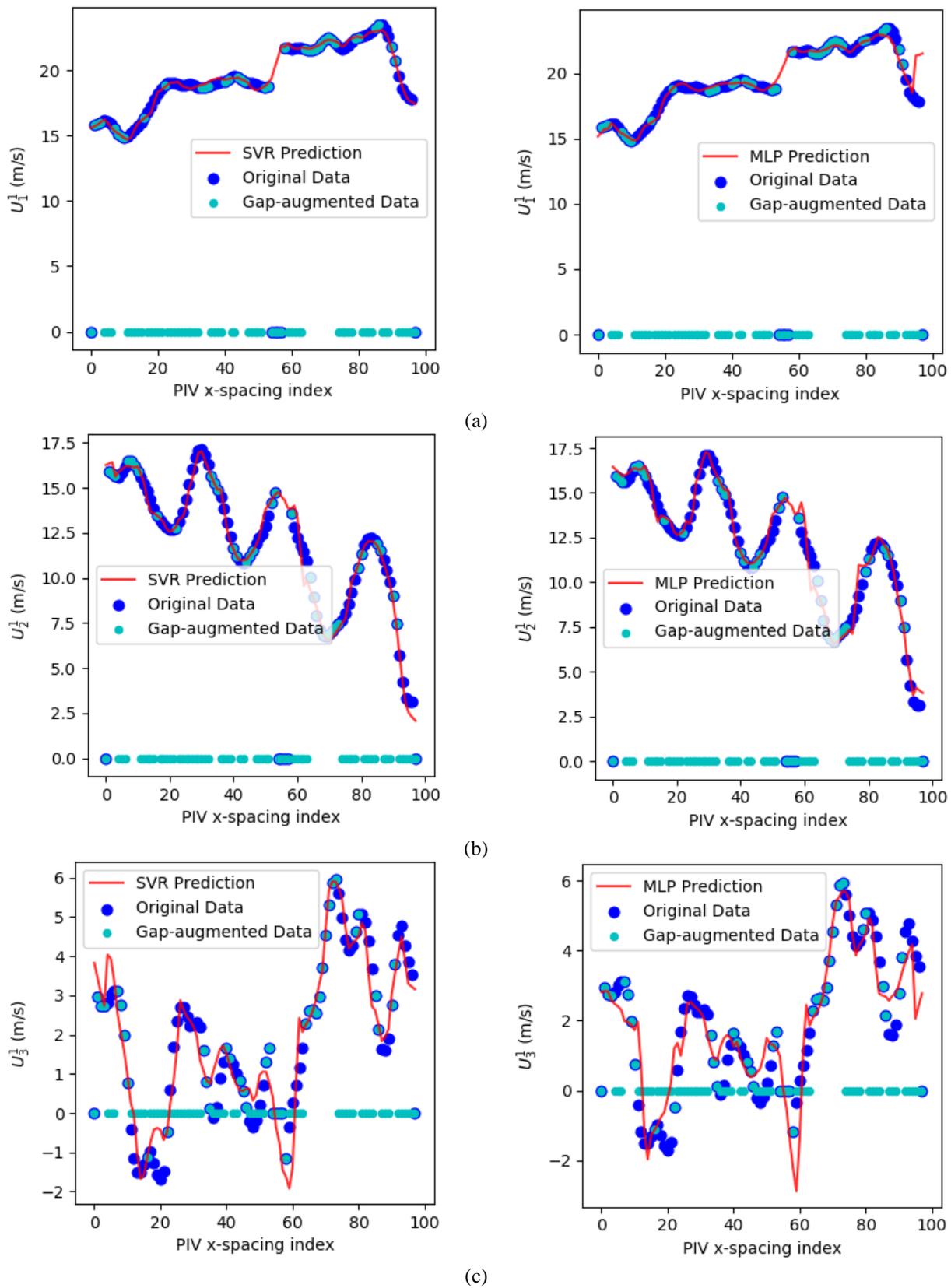

Figure 7. Comparison of SVR and MLP predictions of velocity (solid curves) with the original and artificially gapped velocity (scattered symbols) for (a) x-component, (b) y-component and (c) z-component with respective subscripts 1 to 3.

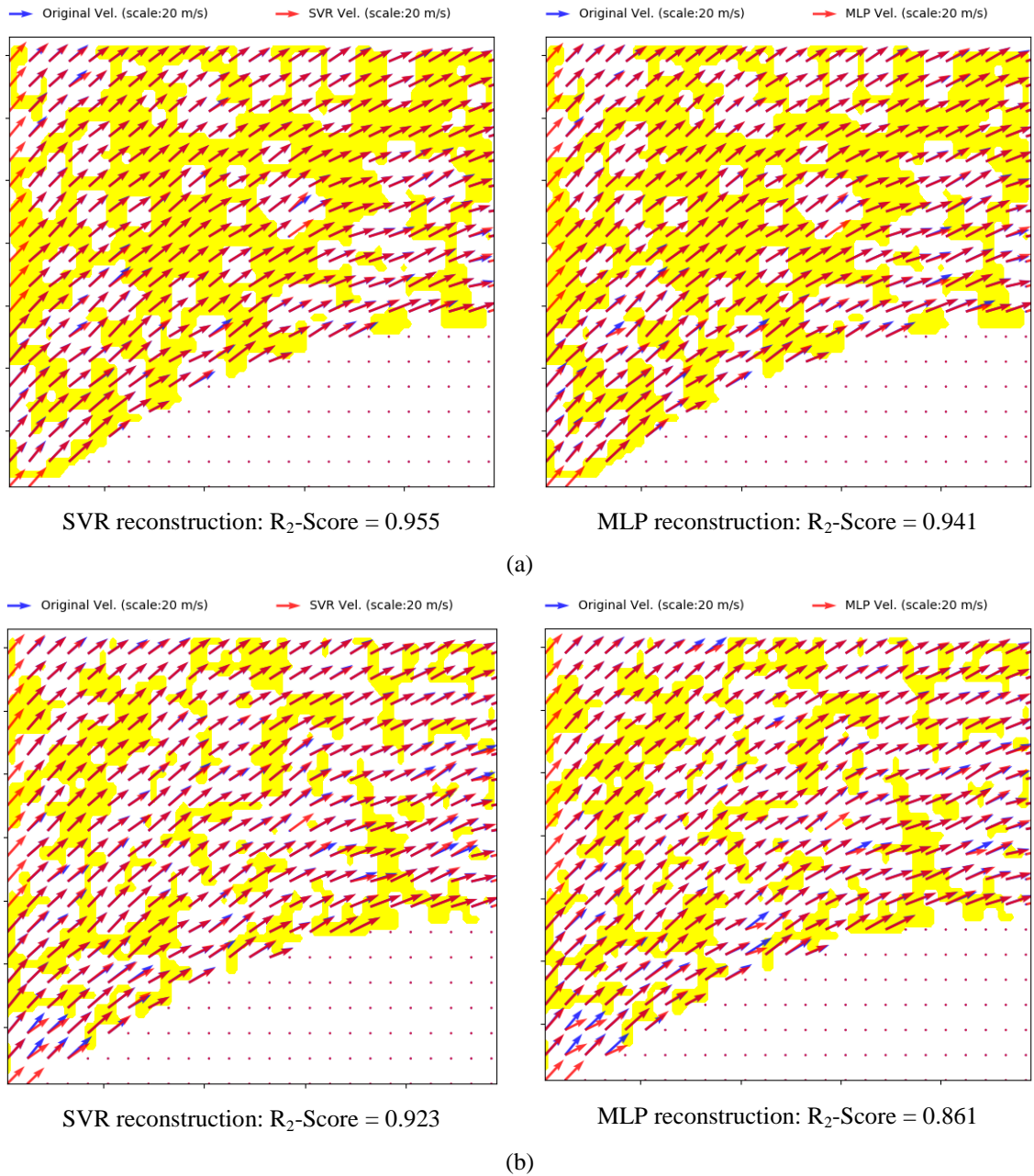

Figure 8. SVR and MLP reconstruction of two-dimensional instantaneous vector maps compared with the true velocity vectors in FOV2 for two different levels of gappiness percentage: (a) 50%; (b) 70%.

Table 6 presents the effect of gap percentage on the accuracy of SVR and MLP regressors in terms of $R_2$-Score for different velocity components. Generally the score for both SVR and MLP approaches decrease with increase of the gap percentage, which is expected. An accuracy reduction rate ($ARR_{\alpha-\beta}$) is defined as the relative decrease of R$_2$-Score by growing the gap percentage from $\alpha\%$ to $\beta\%$, given as:

$$ARR_{\alpha-\beta} = \frac{RS_\alpha - RS_\beta}{RS_\alpha} \times 100\% \qquad (11)$$

where $RS_\alpha$ and $RS_\beta$ are R$_2$-Scores for $\alpha\%$ to $\beta\%$ gaps, respectively. Positive values of accuracy reduction rate correspond to relative reduction of R$_2$-Score by increase of the gap percentage. Comparison of $ARR_{15-60}$ and $ARR_{60-75}$ shows that accuracy reduction is minimum for a 45% growth in gappiness from 15% to 60% while it is significantly higher (particularly for MLP reconstruction) when the gap is further enlarged by just 15%.

*4.5. Hyper-parameter optimization*

The hyper-parameters utilized up to here, were the same for all presented results (as introduced in section 4.1). However, in order to achieve the best predictions by SVR and MLP regressors, their main hyper-parameters are optimized with randomized cross-validation search. The optimal condition is searched such that it maximizes the k-fold cross-validation score. The optimized parameters include the penalty coefficient $C$, the kernel coefficient $\gamma$ and the $\epsilon$ parameter in the SVR model and the number of neurons ($N_1, N_2, N_3$) for MLP with one to three layers.

Table 7 presents the best hyper-parameters for various models averaged for the four FOVs and 100 snapshots. The table also gives the train, test and mean and standard deviation of k-fold cross-validation scores. Overall, the SVR scores are better than values obtained by MLP for all velocity components. Furthermore, both the SVR and MLP models lead to significantly better accuracies in prediction of the in-plane velocity components (with test R$_2$-Scores higher than 0.989 for SVR and 0.97 for two-layer perceptron) in comparison with that of out-of-plane component (with test R$_2$-Score that is at most 0.922 for SVR and 0.898 for two-layer perceptron).

In order to check the state of overfitting in the training process, the difference between R$_2$-Score of train and test procedures are presented in the last column of Table 7. Smallness of train-test score differences confirms the overall absence of overfitting. However, considerably higher values of train-test score difference for the out-of-plane velocity component with respect to the in-plane components indicate that the former has higher potential for overfitting. This could be because of a wavier field for the out-of-plane component. Moreover, comparison of train-test difference for the MLP model for each velocity component shows that using larger number of layers increases that difference and adds to the higher risk of overfitting.

Table 6. Effect of gap percentage on the R$_2$-Score of SVR and MLP models in prediction of separate velocity components.

| ML model | Vel. Comp. | Gap percentage | | | | | Accuracy reduction rate | |
|---|---|---|---|---|---|---|---|---|
| | | 15% | 30% | 45% | 60% | 75% | $ARR_{15-60}$ | $ARR_{60-75}$ |
| SVR | 1 | 0.963 | 0.957 | 0.948 | 0.931 | 0.889 | 3.34 | 4.46 |
| | 2 | 0.979 | 0.977 | 0.974 | 0.964 | 0.937 | 1.52 | 2.81 |
| | 3 | 0.852 | 0.855 | 0.819 | 0.755 | 0.660 | 11.36 | 12.54 |
| MLP | 1 | 0.950 | 0.950 | 0.939 | 0.899 | 0.787 | 5.31 | 12.55 |
| | 2 | 0.970 | 0.971 | 0.966 | 0.955 | 0.877 | 1.57 | 8.11 |
| | 3 | 0.839 | 0.844 | 0.807 | 0.757 | 0.639 | 9.67 | 15.61 |

Table 7. Optimal hyper-parameters for SVR and MLP models including training, testing and cross-validation scores.

| ML Model | Vel. Comp. | Optimal hyper-parameters | | | Evaluation statistics | | | | |
|---|---|---|---|---|---|---|---|---|---|
| | | $C / N_1$ | $\gamma / N_2$ | $\varepsilon / N_3$ | train $R_2$-Score | mean of CV $R_2$-Score | std. dev. of CV $R_2$-Score | test $R_2$-Score | train-test R2-Score difference |
| SVR | 1 | 48.788 | 16.894 | 0.003 | 0.9996 | 0.9945 | 0.0041 | **0.9896** | **0.010** |
| | 2 | 48.153 | 8.725 | 0.014 | 0.9996 | 0.9961 | 0.0020 | **0.9934** | **0.006** |
| | 3 | 28.996 | 4.508 | 0.038 | 0.9893 | 0.9702 | 0.0035 | **0.9221** | **0.067** |
| single-layer perceptron | 1 | 20 | NA | NA | 0.9398 | 0.9395 | 0.0051 | 0.9361 | 0.004 |
| | 2 | 52 | NA | NA | 0.9453 | 0.9443 | 0.0030 | 0.9424 | 0.003 |
| | 3 | 132 | NA | NA | 0.8381 | 0.8322 | 0.0278 | 0.8167 | 0.021 |
| double-layer perceptron | 1 | 47 | 71 | NA | 0.9758 | 0.9751 | 0.0058 | **0.9716** | **0.004** |
| | 2 | 18 | 71 | NA | 0.9850 | 0.9840 | 0.0072 | **0.9809** | **0.004** |
| | 3 | 65 | 75 | NA | 0.9373 | 0.9158 | 0.0142 | **0.8983** | **0.039** |
| triple-layer perceptron | 1 | 37 | 61 | 67 | 0.9836 | 0.9799 | 0.0090 | 0.9688 | 0.015 |
| | 2 | 24 | 64 | 50 | 0.9830 | 0.9819 | 0.0147 | 0.9757 | 0.007 |
| | 3 | 48 | 47 | 62 | 0.9422 | 0.9310 | 0.0223 | 0.8772 | 0.065 |

### *4.6. Statistical and physical analysis of reconstruction errors*

#### *4.6.1. Statistical evaluation based on the velocity components*

In order to analyze the reconstruction error by SVR and MLP for all locations of an instantaneous snapshot, Figure 9 illustrates the boxplot of normalized absolute error ($\Delta U_i^*$, the left frames) as well as absolute error boxplots ($\Delta U_i$, the right frames) calculated based on the following equations:

$$\Delta U_i = \left| U_i^{ML} - U_i^{PIV} \right|, \ i = 1, 2, 3 \tag{12}$$

$$\Delta U_i^* = \frac{\Delta U_i}{U_i^{PIV}}, \ i = 1, 2, 3 \tag{13}$$

where $U_i^{PIV}$ and $U_i^{ML}$ are each of three instantaneous velocity components, measured by PIV or reconstructed by ML regressors, respectively. In each frame, SVM and MLP predictions are compared for different FOVs. The deviation in prediction of the out-of-plane velocity component is one order of magnitude higher than those of the in-plane components in terms of the normalized absolute error. Two reasons are responsible for such trend. First, there is wavier and more complicated field with no general pattern for the third velocity component compared to that of the in-plane components, so it does not follow its neighbouring velocities or any general pattern that might be traced in the snapshot. This causes an extended absolute error between prediction output and target values for the out-of-plane velocity component in comparison with those of planar components (compare the inter-quantile ranges for the frames in the right column of Figure 9 for different velocity components). The second reason is considerably lower values of the third velocity component compared to the in-plane components (see Figure 3 to 5), that enhances the normalized absolute error due to its definition (division by the velocity component in Equation 13). Examination of AE and NAE values as the absolute and relative metrics of the model accuracy shows that in most cases (among different FOVs and velocity components) SVR prediction has resulted in marginally smaller errors. This is the trend in terms of either median, Q1 or Q3 of the boxplots. The normalized absolute errors obtained for FOV1 for the planar velocity components (x and y components), are larger but more scattered in contrast to comparable relative error bands in the second to forth FOVs. The dominance of normalized errors for the y-component velocity field in FOV1 is due to the existence of near-zero values for this velocity component (see part (a) of Figure 4).

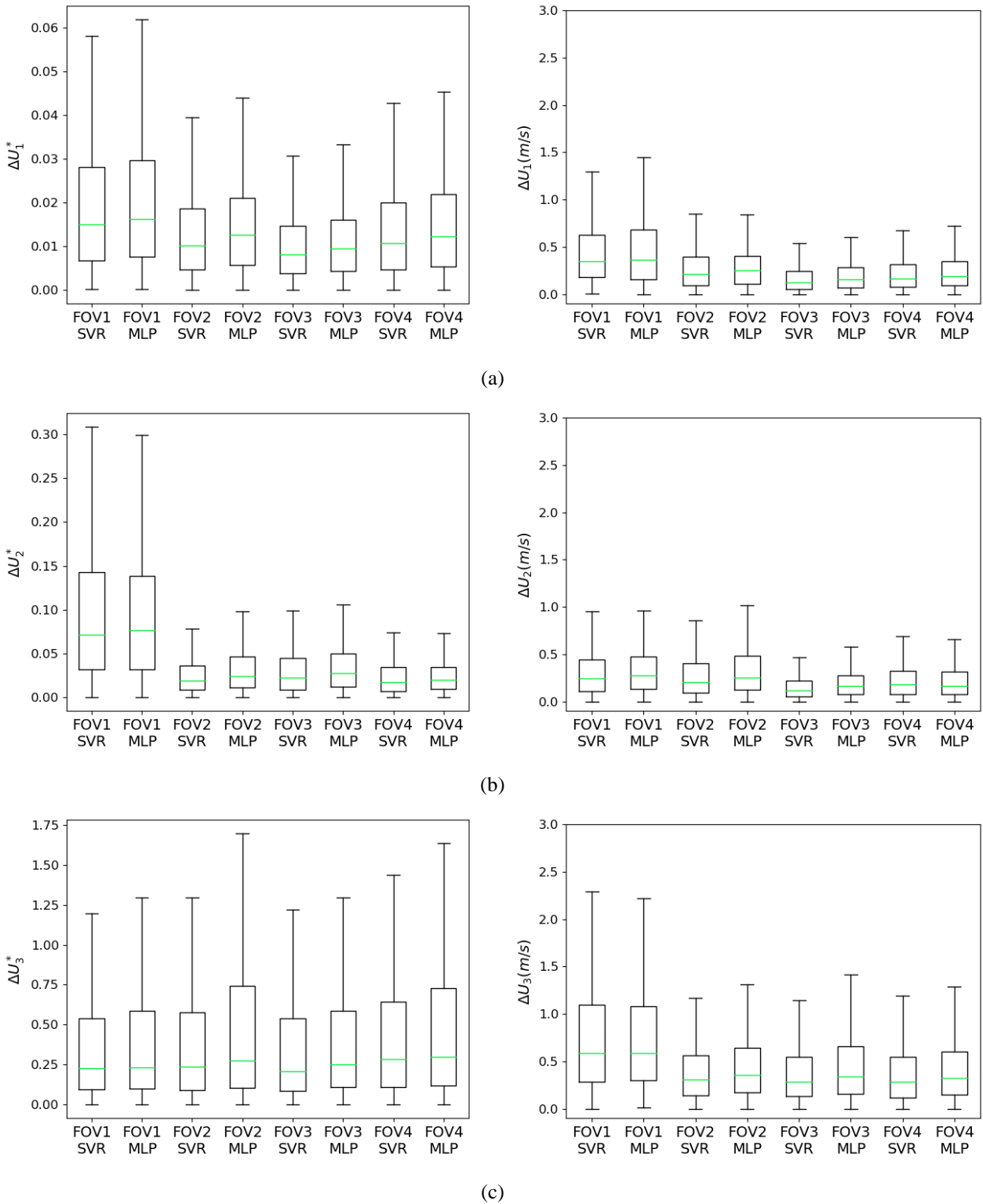

Figure 9. Box plots of prediction error for different velocity components: a) the x-component; b) the y-component; c) the z-component. The left and right frames present the normalized absolute error and the absolute error of the prediction output with respect to the target (PIV measurements), respectively.

*4.6.2. Physical evaluation based on the mass conservation factor*

All the previous sections evaluated the ML model accuracy in terms of statistics calculated based on the velocity components. Consistency of the reconstructed flow field with the original PIV field regarding mass flux conservation is also quite important from fluid mechanics viewpoint. This is evaluated by statistics calculated based on mass conservation factor that we proposed in section 3. Figure 10 illustrates scatter plot of predicted vs. target mass conservation factor for different FOVs. The solid line with a slope of 45º in each plot indicates the exact modeling. More proximity of scattered points to this line indicates more similar ML-based reconstruction of mass flux characteristics in comparison with those obtained from the original PIV flow map as the target. It is obvious that the SVR prediction outperforms the MLP model in this regard.

In order to evaluate the accuracy of $MC$ reproduction by ML models in comparison with that of the original PIV field, different error statistics including mean absolute error ($MAE$), root mean squared error ($RMSE$), median absolute error ($MedAE$) and $R_2$-Score are computed which are presented at the top of each frame in Figure 10. For each statistics, two values are given in the parenthesis corresponding to SVR and MLP models, respectively. Comparison of these pair of values indicates considerably better performance of SVR than MLP in terms of all error metrics. It should be also noticed that $R_2$-Score calculated here based on mass conservation factor is less than those previously evaluated based on in-plane velocity components (Table 7). The reason for this reduction of accuracy is derivative-based definition of $MC$ that significantly intensifies inaccuracy of in-plane velocity component predictions. This decrease is more pronounced in the case of MLP model due to its lower accuracy in reconstruction of velocity components.

Although based on the $R_2$-Scores presented in Figure 10, close accuracies are calculated for different FOVs, other metrics indicate higher absolute errors for the near-rotor regions, namely FOVs 1 and 2. Our previous studies indicated that the flow field at these FOVs involves higher levels of three-dimensionality [40, 44]. This increases the out-of-plane mass flux and consequently elevates the mass conservation factor (as is evident from Figure 10 based on the range of $MC$) and its corresponding absolute error. Figure 11 illustrates box plot of the relative reconstruction error of mass conservation factor ($\Delta MC^*$) for different FOVs and the two ML models. $\Delta MC^*$ is defined as follows:

$$\Delta MC^* = \frac{|MC_{PIV} - MC_{ML}|}{\sigma_{MC,PIV}} \qquad (14)$$

where $\sigma_{MC,PIV}$ is the standard deviation of $MC$ calculated for the test data of each particular FOV. Figure 11 indicates that after normalization of the absolute error of $MC$, the errors for each of the ML models are comparable in different FOVs. In contrast, the median of relative error calculated for MLP is approximately $0.3\sigma_{MC,PIV}$, which is up to 3 times larger than that of the SVR model. This is also valid for the interquartile range that expresses considerably higher variance of $MC$-based error for the MLP model in comparison with that of the SVR.

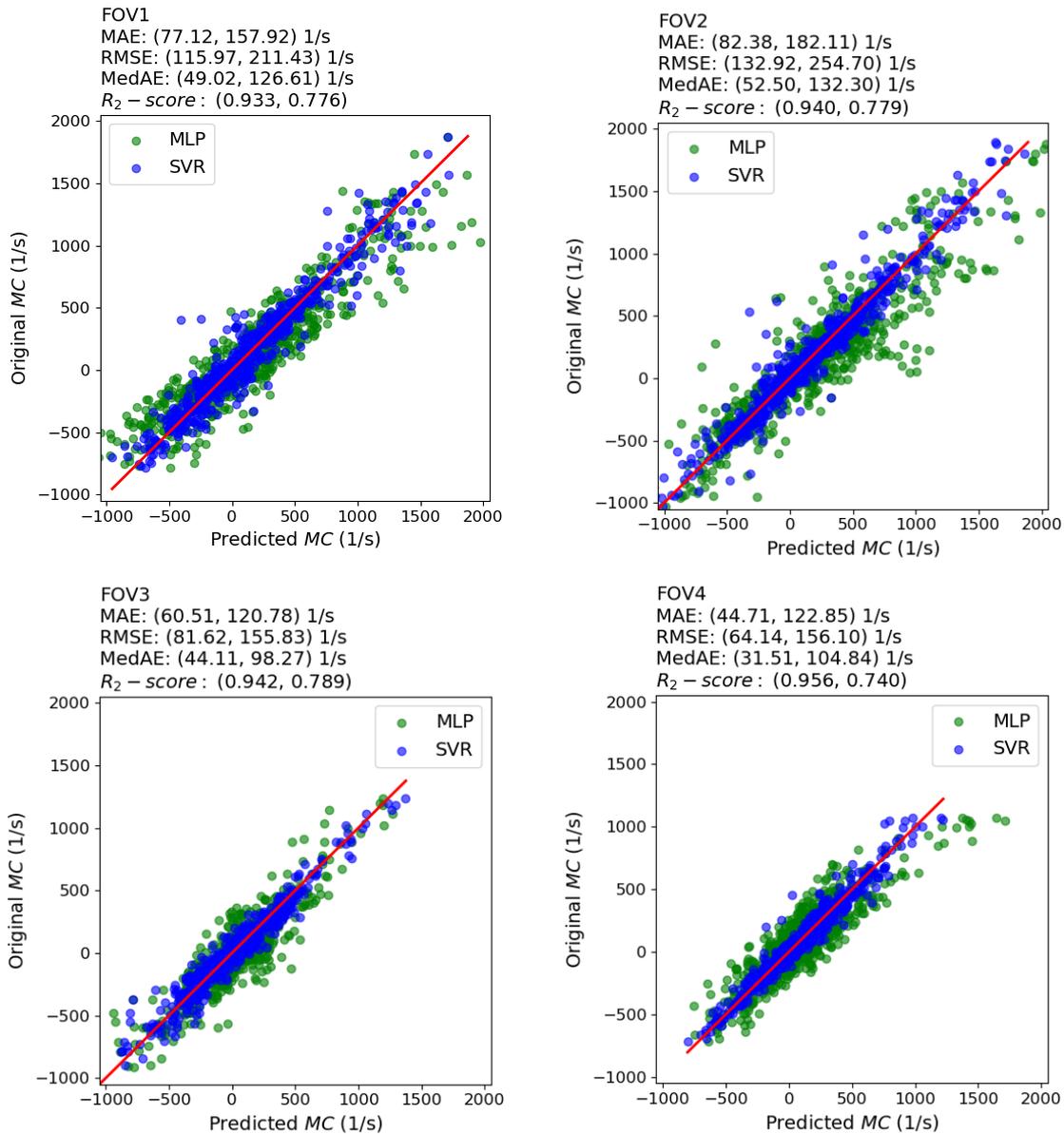

Figure 10. scatter plot of predicted vs. target mass conservation factor for the four FOVs. The FOV number and $MC$ reconstruction accuracy in terms of $MAE$, $RMSE$, $MedAE$ and R$_2$-Score are given at the top of each plot.

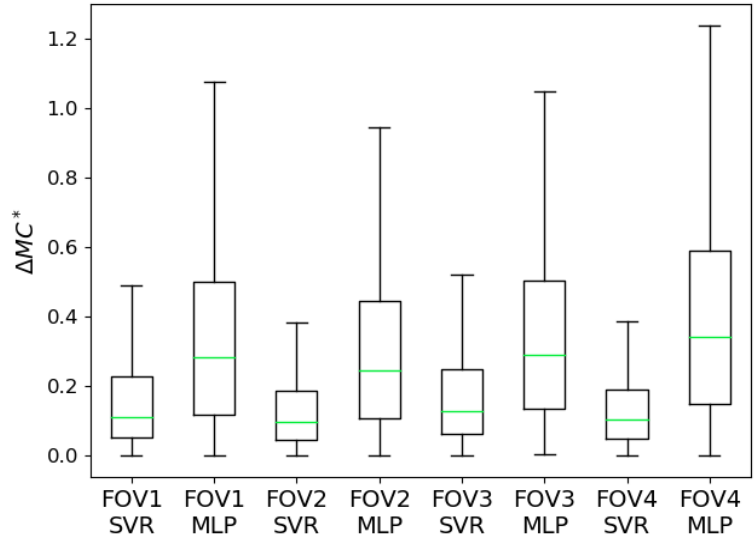

Figure 11. Box plots of normalized absolute error of the mass conservation factor.

## 5. Conclusions

A machine learning approach was established for reconstruction of the measured velocity data of jet/wake flow in the rotor exit region of a centrifugal turbomachine. The SVR and MLP algorithms were chosen for such purpose due to their high potential in handling the non-linear flow phenomena, robustness against overfitting and flexibility in learning procedure. In addition to the conventional coordinate-based features, we proposed practical flow-based features with fundamental fluid mechanics characteristics as the input space in order to enhance the reconstruction accuracy. This opens the way to select correct features that can connect data-based and physics-based models. Train, validation and test data sets were randomly separated after adding numerous artificial gaps with pre-defined size and overall gappiness percentage. The SVR and MLP hyper-parameters were optimally tuned by randomized search, maximizing the k-fold cross-validation score. In addition to velocity-based scores, a metric based on mass conservation was proposed to evaluate the capability of ML models in reconstruction of a physically meaningful flow field regarding fluid dynamics characteristics. The main outcomes of the present study are summarized as follows:

- Implementation of flow-based features that highly correlate with velocity field was quite effective in enhancement of ML-based reconstructions. Examination of different combinations of the generated features showed that a combination including filtered velocity components (based on a Gaussian filter) and the coordinate-based features leads to the highest accuracy in velocity reproduction. Utilization of derivative-based features (filtered vorticity/strain components) was not as effective as the filtered velocity-based features due to their fluctuative behaviour.
- Implementation of different scaling approaches namely normalizer, min-max scaler, standard scaler and robust scaler proved higher performance of the standard scaler due to normal distribution of velocity field data, and competitive function of the robust scaler due to isolating outlier data.
- The proposed approach was able to reproduce extensively clustered missing data by utilizing a larger filter scale. SVR and MLP models satisfactorily reconstructed velocity field for extremely sparse data. This was confirmed after reproduction of reasonable vector plots from these models as compared with the original vector maps by $R_2$-Score as high as 0.94 for the in-plane velocity components and for gappiness levels up to 75%.
- The SVR reconstruction accuracy is slightly superior to that of the MLP outcome, in terms of either median normalized absolute error or $R_2$-Score, for all velocity components and different levels of gappiness.
- After hyper-parameter optimization, the R2 scores for SVR reached to 0.989 for in-plane and 0.922 for out of plane velocity components in comparison with respective scores of 0.97 and 0.898 for MLP.
- The proposed mass conservation metric produced a more distinct advantage for SVR. More specifically, the $MC$-based $R_2$-Scores of SVR and MLP were up to 0.96 and 0.79, respectively and the median of $MC$-based relative error obtained for MLP was up to 3 times larger than that of the SVR model. Furthermore, the $MC$-based $R_2$-Scores obtained for both models was smaller than the velocity-based $R_2$-Scores, indicating more sensitive behaviour of $MC$-based statistics that is because of the derivative-based definition of mass conservation factor.

**References**

[1] M. Raffel, C.E. Willert, F. Scarano, C.J. Kähler, S.T. Wereley, J. Kompenhans, Particle image velocimetry: a practical guide, Springer, 2018.
[2] A. Vlasenko, E.C. Steele, W.A.M. Nimmo-Smith, A physics-enabled flow restoration algorithm for sparse PIV and PTV measurements, Measurement Science and Technology. 26 (2015) 065301.
[3] D. Garcia, A fast all-in-one method for automated post-processing of PIV data, Experiments in fluids. 50 (2011) 1247-1259.
[4] C. Tang, W. Sun, H. He, H. Li, E. Li, Spurious PIV vector detection and correction using a penalized least-squares method with adaptive order differentials, Experiments in Fluids. 58 (2017) 81.
[5] T. Okuno, Y. Sugii, S. Nishio, Image measurement of flow field using physics-based dynamic model, Measurement Science and Technology. 11 (2000) 667.
[6] T. Suzuki, Reduced-order Kalman-filtered hybrid simulation combining particle tracking velocimetry and direct numerical simulation, Journal of Fluid Mechanics. 709 (2012) 249-288.
[7] J. Westerweel, F. Scarano, Universal outlier detection for PIV data, Experiments in fluids. 39 (2005) 1096-1100.
[8] J. Duncan, D. Dabiri, J. Hove, M. Gharib, Universal outlier detection for particle image velocimetry (PIV) and particle tracking velocimetry (PTV) data, Measurement Science and Technology. 21 (2010) 057002.
[9] R.D. Keane, R.J. Adrian, Optimization of particle image velocimeters. I. Double pulsed systems, Measurement science and technology. 1 (1990) 1202.
[10] Y. Lee, H. Yang, Z. Yin, Outlier detection for particle image velocimetry data using a locally estimated noise variance, Measurement Science and Technology. 28 (2017) 035301.
[11] J. Nogueira, A. Lecuona, P. Rodriguez, Data validation, false vectors correction and derived magnitudes calculation on PIV data, Measurement Science and Technology. 8 (1997) 1493.

[12] D.P. Hart, PIV error correction, Experiments in fluids. 29 (2000) 13-22.
[13] C.-S. Pun, A. Susanto, D. Dabiri, Mode-ratio bootstrapping method for PIV outlier correction, Measurement Science and Technology. 18 (2007) 3511.
[14] H. Gunes, U. Rist, Spatial resolution enhancement/smoothing of stereo–particle-image-velocimetry data using proper-orthogonal-decomposition–based and Kriging interpolation methods, Physics of Fluids. 19 (2007) 064101.
[15] R. Everson, L. Sirovich, Karhunen–Loeve procedure for gappy data, JOSA A. 12 (1995) 1657-1664.
[16] D. Venturi, G.E. Karniadakis, Gappy data and reconstruction procedures for flow past a cylinder, Journal of Fluid Mechanics. 519 (2004) 315-336.
[17] S.G. Raben, J.J. Charonko, P.P. Vlachos, Adaptive gappy proper orthogonal decomposition for particle image velocimetry data reconstruction, Measurement Science and Technology. 23 (2012) 025303.
[18] H. Wang, Q. Gao, L. Feng, R. Wei, J. Wang, Proper orthogonal decomposition based outlier correction for PIV data, Experiments in Fluids. 56 (2015) 43.
[19] J. Higham, W. Brevis, C.J. Keylock, A rapid non-iterative proper orthogonal decomposition based outlier detection and correction for PIV data, Measurement Science and Technology. 27 (2016) 125303.
[20] P. Saini, C.M. Arndt, A.M. Steinberg, Development and evaluation of gappy-POD as a data reconstruction technique for noisy PIV measurements in gas turbine combustors, Experiments in Fluids. 57 (2016) 122.
[21] X. Wen, Z. Li, D. Peng, W. Zhou, Y. Liu, Missing data recovery using data fusion of incomplete complementary data sets: A particle image velocimetry application, Physics of Fluids. 31 (2019) 025105.
[22] K. Fukami, K. Fukagata, K. Taira, Assessment of supervised machine learning methods for fluid flows, Theoretical and Computational Fluid Dynamics. (2020) 1-23.
[23] J. Liang, S. Cai, C. Xu, J. Chu, Filtering enhanced tomographic PIV reconstruction based on deep neural networks, IET Cyber-systems and Robotics. 2 (2020) 43-52.
[24] Q. Gao, Q. Li, S. Pan, H. Wang, R. Wei, J. Wang, Particle reconstruction of volumetric particle image velocimetry with strategy of machine learning, arXiv preprint arXiv:1909.07815. (2019).
[25] S. Cai, S. Zhou, C. Xu, Q. Gao, Dense motion estimation of particle images via a convolutional neural network, Experiments in Fluids. 60 (2019) 73.
[26] S. Cai, J. Liang, Q. Gao, C. Xu, R. Wei, Particle image velocimetry based on a deep learning motion estimator, IEEE Transactions on Instrumentation and Measurement. 69 (2019) 3538-3554.
[27] Y. Gim, D.K. Jang, D.K. Sohn, H. Kim, H.S. Ko, Three-dimensional particle tracking velocimetry using shallow neural network for real-time analysis, Experiments in Fluids. 61 (2020) 1-8.
[28] S. Barwey, M. Hassanaly, V. Raman, A. Steinberg, Using machine learning to construct velocity fields from OH-PLIF images, Combustion Science and Technology. (2019) 1-24.
[29] H. Tombul, A.M. Ozbayoglu, M.E. Ozbayoglu, Computational intelligence models for PIV based particle (cuttings) direction and velocity estimation in multi-phase flows, Journal of Petroleum Science and Engineering. 172 (2019) 547-558.
[30] Z. Deng, Y. Chen, Y. Liu, K.C. Kim, Time-resolved turbulent velocity field reconstruction using a long short-term memory (LSTM)-based artificial intelligence framework, Physics of Fluids. 31 (2019) 075108.

[31] X. Jin, S. Laima, W.-L. Chen, H. Li, Time-resolved reconstruction of flow field around a circular cylinder by recurrent neural networks based on non-time-resolved particle image velocimetry measurements, Experiments in Fluids. 61 (2020) 1-23.

[32] A. Giannopoulos, J.-L. Aider, Prediction of the dynamics of a backward-facing step flow using focused time-delay neural networks and particle image velocimetry data-sets, International Journal of Heat and Fluid Flow. 82 (2020) 108533.

[33] B. Liu, J. Tang, H. Huang, X.-Y. Lu, Deep learning methods for super-resolution reconstruction of turbulent flows, Physics of Fluids. 32 (2020) 025105.

[34] Z. Deng, C. He, Y. Liu, K.C. Kim, Super-resolution reconstruction of turbulent velocity fields using a generative adversarial network-based artificial intelligence framework, Physics of Fluids. 31 (2019) 125111.

[35] G. Labonté, Neural network reconstruction of fluid flows from tracer-particle displacements, Experiments in fluids. 30 (2001) 399-409.

[36] J. Pruvost, J. Legrand, P. Legentilhomme, Three-dimensional swirl flow velocity-field reconstruction using a neural network with radial basis functions, Journal of fluids engineering. 123 (2001) 920-927.

[37] L. Casa, P. Krueger, Radial basis function interpolation of unstructured, three-dimensional, volumetric particle tracking velocimetry data, Measurement Science and Technology. 24 (2013) 065304.

[38] S. Al Mamun, C. Lu, B. Jayaraman, Extreme learning machines as encoders for sparse reconstruction, Fluids. 3 (2018) 88.

[39] BS848, Fans for general purposes, part 1: methods for testing performances, British Standard Institution, London. (1997).

[40] G. Akbari, N. Montazerin, On the role of anisotropic turbomachinery flow structures in inter-scale turbulence energy flux as deduced from SPIV measurements, Journal of Turbulence. 14 (2013) 44-70.

[41] M.R. Najjari, N. Montazerin, G. Akbari, On the presence of spectral shortcut in the energy budget of an asymmetric jet–wake flow in a forward-curved centrifugal turbomachine as deduced from SPIV measurements, Journal of Turbulence. 16 (2015) 503-524.

[42] FlowManager software and introduction to PIV instrumentation, Dantec Dynamics A/S, Tonsbakken 18, DK-2740 Skovlunde, Denmark. (2002).

[43] M. Najjari, N. Montazerin, G. Akbari, Statistical PIV data validity for enhancement of velocity driven parameters in turbomachinery jet-wake flow, in: 20th annual international conference on mechanical engineering, Shiraz, Iran, 2012, pp. 16-18.

[44] G. Akbari, N. Montazerin, M. Akbarizadeh, Stereoscopic particle image velocimetry of the flow field in the rotor exit region of a forward-blade centrifugal turbomachine, Proceedings of the Institution of Mechanical Engineers, Part A: Journal of Power and Energy. 226 (2012) 163-181.

[45] S. Liu, C. Meneveau, J. Katz, On the properties of similarity subgrid-scale models as deduced from measurements in a turbulent jet, Journal of Fluid Mechanics. 275 (1994) 83-119.

[46] S.B. Pope, Turbulent Flows, Cambridge University Press, Cambridge, 2000.

[47] O. Gonzalez, A.M. Stuart, A first course in continuum mechanics, Cambridge University Press, 2008.

[48] V. Vapnik, S.E. Golowich, A.J. Smola, Support vector method for function approximation, regression estimation and signal processing, in: Advances in neural information processing systems, 1997, pp. 281-287.

[49] S. Marsland, Machine learning: an algorithmic perspective, CRC press, 2015.

[50] J. Bergstra, Y. Bengio, Random search for hyper-parameter optimization, Journal of machine learning research. 13 (2012) 281-305.
[51] N. Fumo, M.R. Biswas, Regression analysis for prediction of residential energy consumption, Renewable and sustainable energy reviews. 47 (2015) 332-343.
[52] A.K. Varma, D. Mitra, Statistical Feature-Based SVM Wideband Sensing, IEEE Communications Letters. 24 (2019) 581-584.